\documentclass[11pt]{article}
\pdfoutput=1

\usepackage{cancel,slashed}
\usepackage{bbm}
\usepackage{bm}
\usepackage{amsmath,amsfonts,amssymb,mathtools,nicefrac,nccmath,cases}
\usepackage{graphicx}
\usepackage{cite}
\usepackage{skak}

\usepackage{dsfont}
\usepackage{latexsym}
\usepackage{color}
\usepackage[hyperfootnotes=false,linktocpage]{hyperref}
\usepackage{transparent}
\usepackage[hang,flushmargin]{footmisc}
\usepackage{blkarray}
\usepackage{multirow}
\usepackage{empheq}
\usepackage{comment}
\usepackage{wasysym}
\usepackage{tikzsymbols}

\usepackage{caption}
\usepackage{subcaption}
\usepackage[normalem]{ulem}
\usepackage{braket}

\usepackage{hhline}
\usepackage{diagbox}

\newcommand{\bmat}{\left(\begin{array}}
\newcommand{\emat}{\end{array}\right)}

\def\C{\mathbbm{C}}

\def\a {\alpha}
\def\b {\beta}

\def\ov{\overline}

\def\IM{\text{Im}\,}
\def\RE{\text{Re}\,}
\def\ov{\overline}
\def\1{{\bf 1}}
\def\2{{\bf 2}}
\def\3{{\bf 3}}
\def\4{{\bf 4}}
\def\6{{\bf 6}}

\def\targ#1#2{\genfrac{[}{]}{0pt}{}{#1}{#2}}
\def\targ2#1#2{\genfrac{}{}{0pt}{}{#1}{#2}}

\definecolor{mygr}{rgb}{0,0.6,0}
\definecolor{mygrey}{rgb}{0,0.1,0.2}
\definecolor{myblue}{rgb}{0,0.5,0.9}
\definecolor{myblue2}{rgb}{0,0.5,0.5}
\definecolor{myblue3}{rgb}{0,0.7,0.9}
\definecolor{myblue4}{rgb}{0,0.6,0.6}
\definecolor{myorange}{rgb}{1,0.5,0}
\definecolor{mypurple}{rgb}{0.6,0,1}
\definecolor{mygolden}{rgb}{1,0.8,0.2}
\definecolor{mycyan}{rgb}{0,1,1}
\definecolor{mymagenta}{rgb}{1,0,1}
\definecolor{mykiwi}{rgb}{0.8,1,0.5}
\definecolor{mybrown}{cmyk}{0.14, 0.42, 0.56, 0.2}
\definecolor{myturq}{cmyk}{0.99, 0, 0.2, 0.4}
\definecolor{myaubergine2}{cmyk}{0.4, 0.5, 0, 0.1}
\definecolor{myaubergine}{cmyk}{0.6,0.85,0,0}
\definecolor{CycleGreen}{cmyk}{0.52,0,1,0}
\definecolor{CycleBrown}{cmyk}{0, 0.4, 0.9, 0.2}

\DeclareFontFamily{U}{rcjhbltx}{}
\DeclareFontShape{U}{rcjhbltx}{m}{n}{<->rcjhbltx}{}
\DeclareSymbolFont{hebrewletters}{U}{rcjhbltx}{m}{n}

\DeclareMathSymbol{\lamed}{\mathord}{hebrewletters}{108}
\DeclareMathSymbol{\mem}{\mathord}{hebrewletters}{109}
\DeclareMathSymbol{\ayin}{\mathord}{hebrewletters}{96}
\DeclareMathSymbol{\tsadi}{\mathord}{hebrewletters}{118}
\DeclareMathSymbol{\qof}{\mathord}{hebrewletters}{113}
\DeclareMathSymbol{\resh}{\mathord}{hebrewletters}{114}
\DeclareMathSymbol{\pe}{\mathord}{hebrewletters}{112}
\DeclareMathSymbol{\pesofit}{\mathord}{hebrewletters}{80}
\DeclareMathSymbol{\samekh}{\mathord}{hebrewletters}{115}
\DeclareMathSymbol{\tav}{\mathord}{hebrewletters}{116}
\DeclareMathSymbol{\vav}{\mathord}{hebrewletters}{119}
\DeclareMathSymbol{\het}{\mathord}{hebrewletters}{120}
\DeclareMathSymbol{\yod}{\mathord}{hebrewletters}{121}
\DeclareMathSymbol{\zayin}{\mathord}{hebrewletters}{122}
\DeclareMathSymbol{\alephdot}{\mathord}{hebrewletters}{128}
\DeclareMathSymbol{\tsadisofit}{\mathord}{hebrewletters}{90}
\DeclareMathSymbol{\shin}{\mathord}{hebrewletters}{152}

\newcommand{\dd}{\text{d}}

\def\CK {{\cal K}}

\def\be{\begin{equation}}
\def\ee{\end{equation}}
\def\bea{\begin{eqnarray}}
\def\eea{\end{eqnarray}}
\def\bes{\begin{subequations}}
\def\ees{\end{subequations}}

\def\eps{{\epsilon}}
\def\oh{\frac{1}{2}}

\def\re{\mbox{Re}\, }
\def\im{\mbox{Im}\, }
\def\tr{\mbox{Tr}}

\def\om{\omega}

\usepackage{multicol}
\usepackage{float}
\usepackage{caption}
\def\mk {{\mathcal K}}

\def\tr {{\tilde{\rho}}}
\def\p {{\partial}}

\newcommand{\cK}{\mathcal{K}}
\newcommand{\cM}{\mathcal{M}}
\newcommand{\cN}{\mathcal{N}}

\newcommand{\IR}{\mathbb{R}}
\newcommand{\IZ}{\mathbb{Z}}

\renewcommand{\bm}{\boldmath}

\newenvironment{eqn*}{\begin{equation*}\begin{aligned}}{\end{aligned}\end{equation*}\noindent}

\makeatletter
\newsavebox\myboxA
\newsavebox\myboxB
\newlength\mylenA

\newcommand*\xoverline[2][0.75]{%
\sbox{\myboxA}{$\m@th#2$}%
\setbox\myboxB\null
\ht\myboxB=\ht\myboxA%
\dp\myboxB=\dp\myboxA%
\wd\myboxB=#1\wd\myboxA
\sbox\myboxB{$\m@th\overline{\copy\myboxB}$}
\setlength\mylenA{\the\wd\myboxA}
\addtolength\mylenA{-\the\wd\myboxB}%
\ifdim\wd\myboxB<\wd\myboxA%
   \rlap{\hskip 0.5\mylenA\usebox\myboxB}{\usebox\myboxA}%
\else
    \hskip -0.5\mylenA\rlap{\usebox\myboxA}{\hskip 0.5\mylenA\usebox\myboxB}%
\fi}
\makeatother

\begin{document}
\pagestyle{plain}

\makeatletter
\@addtoreset{equation}{section}
\makeatother
\renewcommand{\theequation}{\thesection.\arabic{equation}}

\pagestyle{empty}
\rightline{IFT-UAM/CSIC-23-101}
\vspace{0.5cm}
\begin{center}
\Huge{{New families of scale separated vacua} 
\\[10mm]}
\normalsize{Rafael Carrasco,\footnote{rafael.carrasco@ift.csic.es} Thibaut Coudarchet,\footnote{thibaut.coudarchet@hotmail.fr}  Fernando Marchesano\footnote{fernando.marchesano@csic.es} and David Prieto\footnote{david.prietor@estudiante.uam.es}} \\[12mm]
\small{
Instituto de F\'{\i}sica Te\'orica UAM-CSIC, c/Nicol\'as Cabrera 13-15, 28049 Madrid, Spain 
\\[10mm]} 
\small{\bf Abstract} \\[5mm]
\end{center}
\begin{center}
\begin{minipage}[h]{15.0cm} 

Massive type IIA flux compactifications of the form AdS$_4 \times X_6$, where $X_6$ admits a Calabi--Yau metric and O6-planes wrapping three-cycles, display families of vacua with  parametric scale separation between the compactification scale and the AdS$_4$ radius, generated by an overall rescaling of internal four-form fluxes. For toroidal orbifolds one can perform two T-dualities and map this background to an orientifold of massless type IIA compactified on an SU(3)-structure manifold with fluxes. Via a 4d EFT analysis, we generalise this last  construction and embed it into new branches of supersymmetric and non-supersymmetric vacua with similar features. We apply our results to propose new infinite families of vacua based on elliptic fibrations with metric fluxes. Parametric scale separation is achieved by an asymmetric flux rescaling which, however, in general is not a simple symmetry of the 4d equations of motion. At this level of approximation the vacua are stable but, unlike in the Calabi--Yau case, they display a non-universal mass spectrum of light fields. 

\end{minipage}
\end{center}
\newpage
\setcounter{page}{1}
\pagestyle{plain}
\renewcommand{\thefootnote}{\arabic{footnote}}
\setcounter{footnote}{0}


\tableofcontents


\section{Introduction and summary}
\label{s:intro}

The structure of AdS solutions in string theory plays nowadays a central role in the quest to determine which EFTs are compatible with Quantum Gravity \cite{Vafa:2005ui,Brennan:2017rbf,Palti:2019pca,vanBeest:2021lhn,Grana:2021zvf}. This is mostly due to two current proposals for Swampland criteria known as the AdS instability \cite{Ooguri:2016pdq,Freivogel:2016qwc} and AdS distance \cite{Lust:2019zwm} conjectures, which severely constrain those AdS vacua that can have a holographic CFT dual. The first one proposes that all non-supersymmetric AdS backgrounds are unstable, while the strong version of the second forbids infinite families of supersymmetric AdS vacua with an increasing  separation between the inverse AdS radius and the compactification  scale. 

A class of AdS$_4$ constructions that have become under scrutiny partially due to these conjectures is the so-called DGKT-CFI family of vacua \cite{DeWolfe:2005uu,Camara:2005dc} (see also  \cite{Derendinger:2004jn,Villadoro:2005cu}), which are based on massive type IIA string theory compactified on a smooth Calabi--Yau or a toroidal orbifold $X_6$, with additional $p$-form fluxes threading $X_6$, and an orientifold projection that introduces O6-planes wrapping its three-cycles. It turns out that the vacua equations obtained in this setting lead to infinite families of supersymmetric and non-supersymmetric solutions, generated by an overall rescaling of the internal four-form flux, and which display increasing separation between the AdS$_4$ and compactification scales for larger flux quanta and weaker string coupling.

These constructions, which in the toroidal setting can be supplemented by metric fluxes \cite{Derendinger:2004jn,Villadoro:2005cu,Camara:2005dc}, can be understood as a set of vacua obtained at a 4d EFT level, in the sense that they only solve the 10d equations of motion and Bianchi identities if localised sources like D6-branes and O6-planes are smeared at wavelengths below the compactification scale \cite{Acharya:2006ne}. This fact opened the possibility that solving the actual 10d string theory equations would lead to modifications of the 4d EFT, as it happens in some instances for the K\"ahler potential due to flux backreaction \cite{Font:2019uva}, and this would modify the final set of vacua. A closer look into the 10d description of these backgrounds showed that this does not seem to be the case, at least for the Calabi--Yau setting. Indeed, it was found in \cite{Junghans:2020acz,Marchesano:2020qvg} that the smeared-source approximation can be understood as the leading term of a perturbative expansion of the 10d supergravity equations in either the AdS radius or the string coupling. Such 10d equations were solved up to next-to-leading order, displaying source localisation and an AdS radius and compactification scale in agreement with 4d EFT expectations. Moreover, up to this level of approximation a subset of non-supersymmetric vacua was found to be perturbatively \cite{Marchesano:2019hfb} and (marginally) non-perturbatively \cite{Aharony:2008wz,Narayan:2010em,Marchesano:2021ycx} stable. Extending these results to higher orders in the said perturbative expansion seems a challenging task, since beyond the current level of accuracy one should also take into account $\alpha^\prime$ and string loop effects that take us away from the 10d supergravity description. As a result, it seems involved to provide a more accurate microscopic description of these backgrounds, as it is oftentimes the case with massive type IIA compactifications \cite{Aharony:2010af}.

Since, as of today, the holographic dual of DGKT-CFI vacua has not been found, the debate of the interplay of these constructions with the above Swampland conjectures remains open. One possibility that could make all of them compatible is that an instability is generated at a level of accuracy beyond the one tested so far, which for the supersymmetric branch of vacua would in particular imply that supersymmetry is secretly broken. A more dramatic possibility that would favour the Swampland criteria is that these constructions are inconsistent to begin with, since they mix ingredients like the Romans mass and O6-planes that are separately microscopically well understood, but not combined. To circumvent this second caveat, \cite{Banks:2006hg} proposed to analyse these massive IIA constructions from a dual viewpoint, namely after applying two T-dualities on a $T^2$ factor of a toroidal orbifold, which leads to massless IIA string theory with O6-planes and internal two-form and six-form fluxes, albeit compactified on a twisted torus, non-Calabi--Yau geometry. Subsequently, \cite{Caviezel:2008ik} embedded this T-dual frame in the context of smeared SU(3)-structure compactifications, which allows one to connect with the 4d EFT description of these  orbifolded twisted torus vacua. More recently, \cite{Cribiori:2021djm} implemented the perturbative techniques of \cite{Saracco:2012wc,Junghans:2020acz,Marchesano:2020qvg} in this T-dual setup, finding the expected source-localisation as a correction to the initial SU(3)-structure metric. Moreover, they pointed out the existence of infinite families of vacua related by flux rescalings, that also display parametric scale separation for larger flux quanta. Unlike in the massive IIA frame, for some families of vacua rescaling the fluxes also takes us to a strong 10d string coupling regime, suggesting a family of M-theory flux vacua with scale separation which has nevertheless been challenged in \cite{Collins:2022nux}. 
 
Clearly, testing the properties of AdS string vacua, in particular their consistency, stability and scale separation properties, plays a key role in shaping our vision of the string theory Landscape. However, the general questions under debate will not be answered until we have a global perspective over the whole set of AdS vacua. The aim of this paper is to proceed in the classification of type IIA geometric flux compactifications that lead to AdS$_4$ vacua, addressed from a 4d EFT viewpoint, as already initiated in \cite{Marchesano:2019hfb,Marchesano:2020uqz}. This time we focus on understanding the supersymmetric and non-supersymmetric branches of vacua that include type IIA on (orientifolds of) twisted tori with RR background fluxes but no Romans mass, as in \cite{Banks:2006hg,Caviezel:2008ik,Cribiori:2021djm}, and in particular on those constructions that lead to parametric scale separation upon flux rescaling. 

Our approach leads us to unveil four new branches of 4d vacua, of which one is supersymmetric and three non-supersymmetric, that can be interpreted as a 10d smeared-source compactification on a SU(3)-structure manifold $X_6$ with $p$-form fluxes and non-trivial intrinsic torsion classes, more precisely with the structure of a half-flat manifold. According to the criterion of \cite{Caviezel:2008ik}, any of these branches may contain vacua displaying scale separation. However, we focus on those cases that correspond to backgrounds without Romans mass, as in principle these are easier to analyse microscopically, even beyond the 10d supergravity description. This restriction leads us to consider two subbranches of vacua, one supersymmetric and one non-supersymmetric, that contain the nilmanifold-based compactifications considered in \cite{Caviezel:2008ik,Cribiori:2021djm}, and as such are good candidates to implement the flux scalings of \cite{Cribiori:2021djm} in a more general setting. 

Indeed, one of the main results of this work is to propose new families of AdS$_4$ backgrounds, where scale separation is achieved via a family of vacua generated by flux rescaling, in an analogous manner to the nilmanifold case in \cite{Cribiori:2021djm}. A key observation to arrive to this result is the understanding of such a flux rescaling via the original DGKT-CFI setup. Instead of considering an overall rescaling of all four-form flux quanta as in \cite{DeWolfe:2005uu,Camara:2005dc}, one must implement an asymmetric rescaling in which all K\"ahler moduli grow except for that corresponding to a shrinking $T^2$, so that upon two T-dualities on this $T^2$ one is taken to a large volume compactification. 

With this picture in mind one may implement in new settings some of the flux scalings considered in \cite{Cribiori:2021djm}, that lead to an asymptotic behaviour for the compactification and AdS scale featuring parametric scale separation. An obvious setup in which to do so are Calabi--Yau three-folds that are smooth elliptic fibrations and realise DGKT-like vacua. Upon a Fourier-Mukai transformation on the fibre, these configurations should be taken to SU(3)-structure manifolds with a similar structure of triple intersection numbers and non-vanishing torsion classes, encoded in a metric flux matrix of rank-one. These  topological data are essentially all that one needs to implement our 4d EFT moduli stabilisation techniques, finding similarities but also novelties with respect to the results in \cite{Cribiori:2021djm}. More precisely, we find that in this more general setup the scaling of flux quanta is not a simple symmetry of the equations of motion, so in order to find the family of vacua generated by such a rescaling one must resort to solve them via a perturbative expansion or to a numerical analysis. Nevertheless, the feature of parametric scale separation upon rescaling is still present. Similarly, one finds that the massive spectrum for the stabilised moduli is in general not universal, unlike in DGKT vacua \cite{Marchesano:2019hfb}, and that one only approaches such a spectrum asymptotically. One possible interpretation could be that with these constructions we are testing the dual DGKT-like vacua in small volume regimes where curvature corrections are important, causing such deviations. In any event, since the massive spectrum is not far from the one found in \cite{Marchesano:2019hfb} these vacua are perturbatively stable. In addition, at the smeared-source level of approximation at which we are working, these families of vacua are stable against non-perturbative membrane nucleation process, a statement that holds both for the supersymmetric and non-supersymmetric branches that we consider. 

Some of these results are perhaps not surprising given the T-dual relation of such constructions to DGKT-like vacua with similar properties. It is however important to stress that, in general, T-duality is only expected to hold at the world-sheet perturbative level \cite{Aspinwall:1999ii}, and that any relevant effect that is found for DGKT-CFI vacua beyond that level should be reconsidered for the present set of vacua. This, together with the possibility of finding new infinite families of scale-separated vacua with no T-dual relation to the ones in the literature, lead us to believe that the present construction merit attention on their own, and may account for an important sector of the Landscape of AdS vacua in string theory.

The rest of the paper is organised as follows. In section \ref{s:Tdual} we review the main features of DGKT-CFI vacua from the 4d EFT perspective and discuss four-form flux rescalings different from the ones typically considered in the literature. These shrink some two-tori at substringy areas and are thus better described from a T-dual perspective. In section \ref{s:general} we discuss, from a 4d EFT viewpoint, new branches of type IIA AdS$_4$ vacua that were missed by the analysis of \cite{Marchesano:2020uqz}, and which include the T-duals of DGKT-CFI vacua mentioned above, as follows from their smeared-source 10d description. In section \ref{s:moduli} we apply the techniques of \cite{Marchesano:2020uqz} to these new branches, and describe the vevs of their stabilised moduli in terms of flux quanta, which for factorised geometries translate into scaling symmetries generating  infinite families of vacua with parametric scale separation, reproducing the results of \cite{Cribiori:2021djm}. Section \ref{s:elliptic} applies the same approach to more involved geometries, namely elliptic fibrations with rank-one metric fluxes. In this case the flux scaling symmetry is only approximate, which prompts a perturbative expansion in order to solve the vacua equations. Section \ref{s:stability} studies the perturbative and non-perturbative stability for the set of vacua considered in the previous two sections, finding stability up to our level of approximation. Finally, Appendix \ref{ap:numerical} verifies that the leading-order results on moduli stabilisation and perturbative stability for vacua based on elliptical fibrations are a good approximation to the actual solution, by focusing on two specific models and solving them numerically.


\section{DGKT-CFI vacua and T-duality}
\label{s:Tdual}

Let us consider type IIA string theory compactified on an orientifold of $X_4 \times X_6$ with $X_6$ a compact six-manifold that admits an SU(3)-structure metric defined by the invariant two- and three-form $J$ and $\Omega$, and with volume form $-\frac{1}{6}J^3 = \frac{1}{4} \im \Omega \wedge \re \Omega$. We take the standard  orientifold quotient by $\Omega_p (-1)^{F_L} {\cal R}$ \cite{Blumenhagen:2005mu,Blumenhagen:2006ci,Marchesano:2007de,Ibanez:2012zz,Marchesano:2022qbx}, where $\Omega_p$ is the worldsheet parity-reversal operator,  ${F_L}$ is the spacetime fermion number for left movers, and ${\mathcal R}$ a is an  involution of the metric acting as ${\cal R}(J,\Omega) = -(J, \bar{\Omega})$. This quotient induces a set of O6-planes wrapping three-cycles of $X_6$ which, when backreacted, are presumed to deform the  SU(3) structure to an SU(3)$\times$ SU(3) structure. One however expects that in the limit of smeared sources the initial SU(3)-structure metric is recovered, as it happens in  \cite{Junghans:2020acz,Marchesano:2020qvg,Cribiori:2021djm}. Therefore, in the following we will take such an SU(3)-structure metric as a proxy for the orientifold compactification, following the standard practice in the type IIA moduli stabilisation literature. 

In this framework, one can expand the SU(3)-structure invariant forms $(J,\Omega)$ to define the metric deformations of $X_6$. Following \cite{Gurrieri:2002wz,DAuria:2004kwe,Tomasiello:2005bp,Grana:2005ny,Kashani-Poor:2006ofe} one performs the expansion
\begin{equation}
J_c \equiv B+iJ = \left( b^a + i t^a\right) \om_a \, , 
\end{equation} 
to define the complexified K\"ahler fields $T^a \equiv b^a + i t^a$. For simplicity we will assume that all elements of the two-form expansion basis $\{\om_a\}$ are odd under the action of ${\cal R}$, see e.g. \cite{Marchesano:2020uqz} for the more general case. Similarly, the complex structure deformations are defined via the expansion
\be
\Omega = {\cal Z}^\kappa \alpha_\kappa - {\cal F}_\lambda \beta^\lambda\, ,
\ee
where $(\alpha_\kappa, \beta^\lambda)$ is a symplectic basis of three-forms.\footnote{For Calabi--Yau metrics the set of $p$-forms $\{ \ell_s^{-2} \om_a,  \ell_s^{-3} \alpha_\kappa,  \ell_s^{-3} \beta^\lambda, \ell_s^{-4} \tilde{\om}^b\}$, with $\ell_s = 2\pi \sqrt{\a'}$, corresponds to a basis of harmonic forms that are integrally quantised, and chosen such that $\int \omega_a \wedge \tilde{\omega}^b=\ell_s^6\delta_a^b$ and  $\int \a_\kappa \wedge \b^\lambda=\ell_s^6 \delta_\kappa^\lambda$. For the more general SU(3)-structure case this set may include non-harmonic forms as well, as in the general framework for dimensional reduction on SU(3)-structure manifold developed in \cite{Grana:2005ny,Kashani-Poor:2006ofe}. There the definition of quantised $p$-form becomes more subtle, but it can be made precise in terms of smeared delta forms \cite{Casas:2023wlo}.} The orientifold projection  decomposes this basis into ${\cal R}$-even $(\alpha_K, \beta^\Lambda)$ and ${\cal R}$-odd  $(\beta^K, \alpha_\Lambda)$ 3-forms, and eliminates half of the degrees of freedom of the original complex periods of $\Omega$. 
Then one defines the complexified 3-form $\Omega_c$ as
\begin{equation}
    \Omega_c\equiv C_3+i \im (\mathcal{C}\Omega)\,,
\end{equation}
 where  $\mathcal{C}\equiv e^{-\phi}e^{\frac{1}{2}(K_{\rm cs}-K_K)}$ \cite{Grimm:2004ua}. Here $\phi$ is the 10d dilaton, $K_{\rm cs} \equiv - \log \left(i\ell_s^{-6} \int_{X_6} \ov \Omega \wedge   \Omega \right)$ and $K_K$ is defined as in \eqref{KK}. Finally, the moduli including the complex structure are defined as:
\begin{equation}\label{cpxmoduli}
N^K \equiv \xi^K + i n^K = \ell_s^{-6} \int_{X_6} \Omega_c \wedge \beta^K\, , \qquad  U_{\Lambda} \equiv \xi_\Lambda + i u_\Lambda =  \ell_s^{-6} \int_{X_6} \Omega_c \wedge \alpha_\Lambda\, .
\end{equation}
For simplicity in the following we will collect both sets of moduli $\{N^K, U_\Lambda\}$ into $U^\mu$, introducing a unified basis $\{\alpha_\mu\}= \{\alpha_K,-\beta^\Lambda\}$ and $\{\beta^\mu\}= \{\beta^K,\alpha_\Lambda\}$ and modifying the fluxes accordingly.
In the regime of diluted fluxes, the kinetic terms of all these bulk fields are encoded in a K\"ahler potential of the form $K \equiv K_K + K_Q$, where
\be
K_K \,  \equiv \, -{\rm log} \left(\frac{i}{6} \CK_{abc} (T^a - \bar{T}^a)(T^b - \bar{T}^b)(T^c - \bar{T}^c) \right) \, = \,  -{\rm log} \left(\frac{4}{3} \cK\right) \, ,
\label{KK}
\ee
and 
\begin{equation}\label{KQ}
 K_Q \equiv -2 \log \left( \frac{1}{4} \RE({\cal C}{\cal Z}^\Lambda) \IM({\cal C} {\cal F}_\Lambda) - \frac{1}{4} \IM({\cal C} {\cal Z}^K) \RE({\cal C} {\cal F}_K) \right) = 4\phi_4 \, .
\end{equation}
Here ${\cal K}_{abc} \equiv - \ell_s^{-6} \int_{X_6} \omega_a \wedge \omega_b \wedge \omega_c$ are triple intersection numbers, $\cK \equiv \cK_{abc} t^at^bt^c = 6 {\rm Vol}_{X_6}$ and $\phi_4 = \phi - \oh \log {\rm Vol}_{X_6}$ is the 4d dilaton.

An interesting setup is the case where the smeared metric for $X_6$ is Calabi--Yau as then, by including RR and NSNS three-form fluxes, one can apply the analysis of \cite{DeWolfe:2005uu,Marchesano:2019hfb} to generate infinite families of AdS$_4$ vacua. We work with the democratic formulation of type IIA supergravity \cite{Bergshoeff:2001pv},  in which all RR potentials are grouped in a polyform ${\bf C} = C_1 + C_3 + C_5 + C_7 + C_9$, and so are their gauge invariant field strengths
\be
{\bf G} \,=\, \dd_H{\bf C} + e^{B} \wedge {\bf \bar{G}} \, ,
\label{bfG}
\ee
with $H$ the three-form NSNS flux, $\dd_H \equiv (\dd - H \wedge)$ is the $H$-twisted differential  and ${\bf \bar{G}}$ a formal sum of closed $p$-forms on $X_6$. 

The fluxes enter the perturbative superpotential $W\equiv W_{\rm RR} + W_{\rm NS}$, where $W_{\rm RR}$ involves K\"ahler fields and RR fluxes \cite{Taylor:1999ii}
\begin{equation} \label{WRR}
\ell_s W_{\rm RR} = - \frac{1}{\ell_s^5} \int_{X_6} {\ov{\bf G}} \wedge e^{J_c}= e_0 + e_a T^a + \frac{1}{2} {\cal K}_{abc} m^a T^b T^c + \frac{m}{6} {\cal K}_{abc} T^a T^b T^c\,,
\end{equation}
while $W_{\rm NS}$ contains the complex structure fields and $H$-flux quanta
\begin{equation} \label{WNS}
\ell_s W_{\rm NS} = -\frac{1}{\ell_s^5} \int_{X_6} \Omega_c \wedge H = h_K N^K  + h^\Lambda U_{\Lambda} \equiv h_\mu U^\mu\, .
\end{equation}
Here $e_0, e_a, m^a, m, h_\mu$ are all integers, defined following the same conventions as in \cite{Marchesano:2021ycx}
\begin{equation}
\begin{gathered}
m \, = \,  \ell_s G_0\, ,  \qquad  m^a\, =\, \frac{1}{\ell_s^5} \int_{X_6} \bar{G}_2 \wedge \tilde \omega^a\, , \qquad  e_a\, =\, - \frac{1}{\ell_s^5} \int_{X_6} \bar{G}_4 \wedge \omega_a \, ,\\
\quad e_0 \, =\, - \frac{1}{\ell_s^5} \int_{X_6} \bar{G}_6 \,,\qquad 
h_\mu=\frac{1}{\ell_s^5}\int_{X_6} H\wedge \alpha_\mu\,.
\end{gathered}
\label{RRfluxes}
\end{equation}
These coefficients of the superpotential are redefined when taking into account curvature corrections, see e.g. 
 \cite{Marchesano:2021ycx}.

Let us now focus on the case where $X_6 =(T^2)^3/\Gamma$ with $\Gamma \subset SU(3)$ a discrete orbifold group, where one can apply the results of \cite{DeWolfe:2005uu,Camara:2005dc} (see also \cite{Derendinger:2004jn,Villadoro:2005cu}) to provide explicit expressions for the untwisted moduli vevs in terms the flux quanta. For instance, in the case where $\Gamma = \IZ_2 \times \IZ_2$ the K\"ahler potentials for the untwisted sector take the simplified form $K_K = - \log \left( 2i (T^1 - \bar{T}^1)(T^2 - \bar{T}^2)(T^3 - \bar{T}^3) \right)$ and $K_Q = -\log \left(u^0 u^1u^2u^3\right)$, with $u^i \equiv \im U^i$,  while the corresponding superpotentials take the form \eqref{WRR} and \eqref{WNS} with ${\cal K}_{123} = 2$ and ${\cal K}_{iij}=0$, $\forall i,j$. In this case, the vacua analysis of the 4d F-term scalar potential yields the following relation between K\"ahler moduli and fluxes:
\be
t^a \propto \frac{\sqrt{|\hat{e}_1\hat{e}_2 \hat{e}_3}|}{|\hat{e}_a|}\, ,\quad \text{where} \quad  \hat{e}_a \equiv e_a  - \oh \frac{{\cal K}_{abc}m^bm^c}{m}\, .
\ee

From here, one can explore different branches of vacua by varying the flux quanta. The simplest one correspond to the rescaling
\be
\hat{e}_a \mapsto n^{2}  \hat{e}_a  \quad \implies \quad t^a \mapsto n t^a\, ,
\label{basicsc}
\ee
for $ n \in \mathbb{N}$, which corresponds to an overall rescaling of the volume of $X_6$. This is the scaling used in \cite{DeWolfe:2005uu,Camara:2005dc} to construct an infinite family of scale-separated vacua, which can be implemented in a general Calabi--Yau. In the present setup one can however consider richer scalings, that depend differently on each flux quanta. For instance:
\be
\hat{e}_1 \mapsto n^{2r} \hat{e}_1, \quad  \hat{e}_{i} \mapsto n^{r+s} \hat{e}_{i}\, ,   \quad \implies \quad t^{1} \mapsto n^{s} t^1 ,  \quad t^i \mapsto n^r t^i\, , \qquad   i=2,3\, .
\label{newsc}
\ee
For $r=s=1$, one recovers \eqref{basicsc}, but other values for $r,s$ lead to other branches of vacua. One may then generalise the analysis of \cite{DeWolfe:2005uu,Camara:2005dc,Marchesano:2019hfb} to arrive at the following scaling dependence for the remaining moduli
\be
u^\mu \mapsto n^{2r+s} u^\mu , \quad g_s \mapsto n^{-r-\oh s} g_s\, , \quad e^{\phi_4} \mapsto n^{-2r - s} e^{\phi_4}\, .
\label{newsc2}
\ee
To address the separation of scale, we define the AdS radius $R_{\rm AdS}$ as 
\be
(R_{\rm AdS}M_{\rm P})^{-1}\equiv\sqrt{|\Lambda|}\, ,
\ee
where $\Lambda$ the 4d vacuum energy in 4d Planck mass units $M_{\rm P}$. In the current setup we have \cite{Marchesano:2019hfb}
\begin{equation}
\sqrt{|\Lambda|}\sim\frac{(t^1t^2t^3)^{\frac{1}{2}}}{(u^\mu)^2}\quad\implies\quad  (R_{\rm AdS}M_{\rm P})^{-1}\sim n^{-3r-\frac{3}{2}s}\, .
\end{equation}
This last scaling is to be compared with that of the KK or winding mode scale $M_{\rm KK}$ and $M_{\rm w}$ 
\begin{equation}
\frac{M_{\rm KK}}{M_{\rm P}}\equiv \frac{e^{\phi_4}}{\max\sqrt{t^i}}\sim e^{\phi_4}n^{\min\{-\frac{r}{2},-\frac{s}{2}\}}\, ,\qquad \frac{M_{\rm w}}{M_{\rm P}}\equiv e^{\phi_4}\min\sqrt{t^i}\sim e^{\phi_4}n^{\min\{\frac{r}{2},\frac{s}{2}\}}\, .
\end{equation}
We thus find
\be
 R_{\rm AdS} M_{\rm KK} \sim n^{{\rm min}\{r, \oh (r+s)\} }\, , \ \  R_{\rm AdS} M_{\rm w} \sim n^{{\rm min}\{(r+s), \oh (3r+s)\} }\, ,
\label{scalingsep}
\ee
obtaining that if $r>0$ and $r+s>0$ there is an asymptotic separation of scales. This is true even when $s < 0$, which is a possibility as long as the flux $m^1$ vanishes. Choosing $s<0$ takes us to a small radius limit in the torus factor $(T^2)_1$ associated with $t^1$. In that case, it is more sensible to describe the compactification in a T-dual frame, for instance after two T-dualities on $(T^2)_1$, which flips the sign of $s$ in \eqref{newsc}. This double T-duality is precisely the frame considered in \cite{Banks:2006hg,Caviezel:2008ik,Cribiori:2021djm} to describe a family of vacua in a massless type IIA compactification, which explains the similarity of the scalings above with those obtained in \cite{Cribiori:2021djm}.\footnote{Analogous scalings have been considered in compactifications of massive type IIA in toroidal $G_2$ orientifolds \cite{Farakos:2023nms}.} The following sections aim  to generalise such scalings for non-toroidal setups.

A simple approach for such a generalisation is to describe the two T-dualities at the level of the 4d effective theory. This can be implemented by performing the following change of variables
\be
T^1 \mapsto  -\frac{1}{T^1}\, ,
\label{Tinv}
\ee
in terms of which the K\"ahler potential transforms as
\be
K_K \mapsto K_K + \log |T^1|^2\, , \qquad K_Q \mapsto K_Q\, ,
\label{Ktrans}
\ee
where we have used that the 4d dilaton is invariant under T-duality. The shift in the K\"ahler potential can be absorbed in a K\"ahler transformation
\be
K_K \mapsto K_K -F - \bar{F}\, ,\qquad W \mapsto e^{F} W\, ,
\ee 
with $F \equiv\log T^1$. After performing it, we recover the previous K\"ahler potential and a superpotential of the form 
\begin{subequations}
\label{newW}
\begin{align}
\ell_s W_{\rm RR}  &= T^1\left(e_0 + e_2T^2+e_3T^3 \right) - e_1 -2m^2T^3 - 2m^3 T^2 - 2mT^2T^3\, , 
\label{newWRR} \\
\ell_s W_{\rm NS} & =  h_\mu U^\mu T^1\, .
\label{newWNS}
\end{align}
\end{subequations}
One can interpret the new superpotential in terms of the set of fluxes that one obtains after two T-dualities. Notice that the would-be cubic term $m^1T^1T^2T^3$ is missing due to consistency with an exact scaling of the form \eqref{newsc} for $s<0$.  The lack of cubic term in \eqref{newWRR} signals the absence of Romans mass $G_0$ in the new T-dual frame, while the bilinear structure of \eqref{newWNS} indicates that the former $H$-flux has been traded for a metric flux \cite{Villadoro:2005cu,Derendinger:2004jn,Camara:2005dc}. 

At the perturbative level, the F-term potential derived from \eqref{newW} is equivalent to the previous one with superpotential \eqref{WRR} + \eqref{WNS}. Therefore the 4d vacua analysis leading to the scalings \eqref{scalingsep} works mutatis mutandis as before. However, if in the current frame we want to stay in the regime of large compactification volumes where the 10d supergravity approximation can be trusted, one is prompted to take $s<0$ in the original setting \eqref{newsc}, which means that the standard DGKT-CFI scaling \eqref{basicsc} cannot be implemented. Still, a negative $s$ allows for  parametric scale separation in \eqref{scalingsep}, which translates into an infinite family of scale separated vacua at large volume in the T-dual setup. The existence of such a family was  pointed out in \cite{Cribiori:2021djm},  precisely in this toroidal setup. 

Achieving scale separation with the superpotential \eqref{newW} brings us to the question of how this family of vacua compares to those obtained in \cite{Marchesano:2020uqz}, where a general analysis of type IIA AdS$_4$ vacua with $p$-form and metric fluxes was performed, assuming a specific F-term pattern and obtaining no parametric scale separation.  As we will see in the next section, the AdS$_4$ vacua analysed in \cite{Caviezel:2008ik,Cribiori:2021djm} belong to a branch of vacua which was missed in the classification performed in \cite{Marchesano:2020uqz}, but that one can characterise using the same techniques. This characterisation will allow us to generalise the construction in \cite{Banks:2006hg,Caviezel:2008ik,Cribiori:2021djm} to more involved settings, unveiling new branches of AdS$_4$ orientifold vacua that in principle can realise parametric scale separation. Since these new constructions necessarily feature metric fluxes, it is not so obvious how to identify the set of compact manifolds $X_6$ that they correspond to. However, the above discussion gives us a hint on the simplest geometries where similar scalings to the above should be realised: Calabi--Yaus with an $SL(2,\IZ)$ perturbative duality on a K\"ahler modulus. These geometries include toroidal orbifolds  and smooth elliptic fibrations. In section \ref{s:elliptic} we will see how to use the latter to propose new families of  AdS$_4$ vacua with parametric scale separation. 


\section{General class of 4d vacua}
\label{s:general}

In this section, we review the 4d approach taken in \cite{Marchesano:2020uqz} to classify type IIA AdS$_4$ orientifold vacua, and point out branches of solutions that were missed in the original classification. As we will show, the supersymmetric family of AdS$_4$ vacua with scale separation discussed in \cite{Cribiori:2021djm} is part of these new branches, as well as its non-supersymmetric counterpart. The framework of \cite{Marchesano:2020uqz} allows us to describe the new branches in a general manner, without specifying the underlying SU(3)-structure manifold, but providing certain constraints for its torsion classes which we discuss in the (smeared) 10d interpretation of these vacua. In subsequent sections we apply this general description to propose new families of scale-separated vacua.

\subsection{4d analysis}

Let us briefly review the results of \cite{Marchesano:2020uqz}, to see how new branches of vacua arise from the 4d analysis developed therein. The working assumption of \cite[section 4]{Marchesano:2020uqz} is a 4d K\"ahler potential $K = K_K + K_Q$ that is the sum of \eqref{KK} and \eqref{KQ}, and a superpotential $W= W_{\rm RR} + W_{\rm NS}$, with $W_{\rm RR}$ given by \eqref{WRR} and $W_{\rm NS}$ by
\begin{equation} \label{WNS2}
\ell_s W_{\rm NS} =  h_\mu  U^\mu + f_{a\mu} T^a U^\mu\, ,
\end{equation}
where $f_{a\mu} \in \IZ$ represent metric fluxes. In practice, they are defined in terms of the action of the twisted external derivative  over the  basis  introduced in the previous section
\begin{equation}
    \ell_s\dd \om_a = -f_{a \mu}\, \beta^\mu \, , \qquad  \ell_s\dd \alpha_\mu = -f_{a \mu} \, \tilde\om^a\, ,
    \label{deffamu}
\end{equation}
in agreement with the general scheme for dimensional reduction in SU(3)-structure manifolds \cite{Grana:2005ny,Kashani-Poor:2006ofe}. We refer to \cite[Appendix A]{Marchesano:2020uqz} for more details regarding the action of the metric fluxes. When working with the smeared approximation in this setup, the Bianchi identity for the RR two-form flux threading $X_6$ reads
\be
 \left[ m^a f_{a\mu} + m h_\mu + m f_{a\mu}b^a \right] \b^\mu +  N_\a \delta^\a_{\rm D6} - 4\delta_{\rm O6} = 0 \, ,
\label{BIG2}
\ee
where $\delta_{\rm O6/D6}$ are smeared delta forms of the three-cycles wrapped by D6-branes and O6-planes, see \cite{Casas:2023wlo} for a precise definition. Therefore, the tadpole cancellation condition puts constraints on the available values of the flux quanta $m$, $m^a$, $h_\mu$ and $f_{a\mu}$, but not on the rest. 

From the above superpotential and K\"ahler potential, it follows that the F-terms read
\bes
\begin{align}
F_a \equiv D_{T^a} W=&\left[\rho_a+\frac{\p_a K}{2}\left(t^b\rho_b+u^\mu\rho_\mu+\frac{1}{6}\mk\tr\right)\right]\nonumber\\ +&i\left[\mk_{ab}\tr^b+\rho_{a\mu}u^\mu-\frac{\p_a K}{2}\left(\rho_0-t^bu^\mu\rho_{b\mu}-\frac{1}{2}\mk_b\tr^b\right)\right]\, \label{eq: F-Ta}\, ,\\
F_\mu \equiv D_{U^\mu} W=&\left[\rho_\mu+\frac{\p_\mu K}{2}\left(t^a\rho_a+u^\nu\rho_\nu-\frac{1}{6}\mk\tr\right)\right]\nonumber \\ +&i\left[t^a\rho_{a\mu}-\frac{\p_\mu K}{2}\left(\rho_0-t^au^\nu\rho_{a\nu}-\frac{1}{2}\mk_b\tr^b\right)\right]\, ,
\label{eq: F-Umu}
\end{align} 
\ees
where we have made use of the shift symmetries of the K\"ahler potential to define $\p_a K \equiv \p_{t^a} K$, $\p_\mu K \equiv \p_{u^\mu} K$ and $\CK_a \equiv \CK_{abc} t^bt^c$, $\CK_{ab} \equiv \CK_{abc}t^c$. From here one obtains the F-term potential 
\begin{align}
   \kappa_4^2 V_F =\, &e^K\left[4\rho_0^2+g^{ab}\rho_a\rho_b+\frac{4\mathcal{K}^2}{9}g_{ab}\tilde{\rho}^a\tilde{\rho}^b+\frac{\mathcal{K}^2}{9}\tilde{\rho}^2\right.\nonumber\\
    + &\left.c^{\mu\nu}\rho_\mu\rho_\nu+\left(\tilde{c}^{\mu\nu}t^at^b+g^{ab}u^\mu u^\nu \right)\rho_{a\mu}\rho_{b\nu}-\frac{4\mathcal{K}}{3}u^\nu \tilde{\rho}^a\rho_{a\nu}+\frac{4\mathcal{K}}{3}u^\nu \tilde{\rho}\rho_\nu\right]\, ,
    \label{eq:potentialgeom}
    \end{align}
where the relevant field space metrics are 
\be
g_{ab} \equiv \frac{1}{4} \partial_{t^a} \partial_{t^b} K_K = \frac{3}{2\CK} \left(\frac{3}{2} \frac{\CK_a\CK_b}{\CK} - \CK_{ab}\right) , \quad c_{\mu\nu} \equiv \frac{1}{4} \partial_{u^\mu}\partial_{u^\nu} K_Q = \frac{1}{4G}\left( \frac{G_\mu G_\nu}{G}-G_{\mu\nu}\right) ,
\ee
with $G\equiv e^{-K_Q}$, $G_\mu\equiv\partial_{u^\mu} G$ and $G_{\mu\nu}\equiv\partial_{u^\mu} \partial_{u^\nu} G$. 
Upper indices denote their inverses and $\tilde{c}^{\mu\nu}\equiv c^{\mu\nu}-4u^\mu u^\nu $. Finally, we have defined the flux-axion polynomials
\bes
\label{RRrhosgeom}
\begin{align}
  \ell_s  \rho_0&=e_0+e_ab^a+\frac{1}{2}\mathcal{K}_{abc}m^ab^bb^c+\frac{m}{6}\mathcal{K}_{abc}b^ab^bb^c+ \ell_s \rho_\mu\xi^\mu\, , \label{eq:rho0g}\\
 \ell_s   \rho_a&=e_a+\mathcal{K}_{abc}m^bb^c+\frac{m}{2}\mathcal{K}_{abc}b^bb^c+ f_{a\mu}\xi^\mu \, ,  \label{eq:rho_ag}\\
  \ell_s  \tilde{\rho}^a&=m^a+m b^a \, ,  \label{eq:rho^ag}\\
 \ell_s   \tilde{\rho}&=m \, ,   \label{eq:rhomg}\\
 \ell_s    \rho_\mu&=h_\mu+f_{a\mu}b^a \, , \label{eq:rhomu}\\
 \ell_s   \rho_{a\mu}&=f_{a\mu} \, ,  \label{eq: ho_ak}
\end{align}   
\ees 
where $\xi^\mu \equiv \re U^\mu$. Notice that, up to now, we have not made any assumption on the background fluxes, except that they correspond to a geometric compactification. As before, we have assumed that there are no two-forms $\varpi_\a$ that are even under the orientifold action, which in particular prevents a D-term contribution $V_D$ to the potential. This assumption will not change our results, since as shown in \cite{Marchesano:2020uqz} $V_D$ vanishes at the class of vacua that we will consider.

The flux potential \eqref{eq:potentialgeom} has the bilinear structure $\kappa_4^2\, V_F  = \, Z^{\cal AB} {\rho}_{\cal A} {\rho}_{\cal B}$ stressed in \cite{Bielleman:2015ina,Carta:2016ynn,Herraez:2018vae}, where the matrix entries $Z^{\cal AB}$ only depend on the saxions $\{t^a, n^\mu \}$, while the ${\rho}_{\cal A}$ displayed in \eqref{RRrhosgeom} are gauge-invariant combinations of flux quanta and the axions $\{b^a, \xi^\mu\}$. This allows to implement the techniques used in \cite{Escobar:2018tiu,Escobar:2018rna,Marchesano:2019hfb,Marchesano:2021gyv} to perform a systematic search for vacua. This approach was applied in \cite{Marchesano:2020uqz}, together with the assumption of on-shell F-terms of the form
\be
\langle D_{T^a} W, D_{U^\mu}W \rangle_{\rm vac}=\langle\lambda_K \partial_{T^a} K,\lambda_Q \partial_{U^\mu} K \rangle_{\rm vac}\, ,
\label{solsfmax}
\ee
with  $\lambda_K\ell_s , \lambda_Q\ell_s \in \C$, which includes all supersymmetric vacua. Translated to the language of flux-axion polynomials, this F-term constraint results in the following on-shell relations
\bes
\label{proprhog}
\begin{align}
   \rho_a & = \ell_s^{-1} {\mathcal P}\, \partial_a K \label{eq: f-term prop rho_a geom}\, ,\\
       \mathcal{K}_{ab}\tilde{\rho}^b+\rho_{a\mu}u^\mu & =  \ell_s^{-1} {\mathcal Q}\, \partial_a K \label{eq: f-term prop rho^a geom}\, ,\\
    \rho_\mu & =  \ell_s^{-1}\cM\, \partial_\mu K\, ,  \label{eq: f-term prop rhomu geom} \\
   t^a\rho_{a\mu } & =  \ell_s^{-1} \cN\, \partial_\mu K\, , \label{eq: f-term prop rhoak geom}
\end{align}
\ees
where ${\mathcal P}$, ${\mathcal Q}$, $\cM$, $\cN$ are real functions of the 4d light fields. The resulting F-terms are given by 
\bes
\label{eq: F-terms + ansatz}
\begin{align}
\ell_s F_a =&\left[\mathcal{P}+\frac{1}{2}(-3\mathcal{P}-4\mathcal{M}+\frac{1}{6}\cK \ell_s\tilde{\rho})\right]\partial_a K+i\left[\mathcal{Q}-\frac{1}{2}(\ell_s\rho_0+2\mathcal{N}+\frac{3}{2}\mathcal{Q})\right]\partial_a K\, , \label{eq: F-Ta + ansatz} \\
\ell_s F_\mu =&\left[\mathcal{M}+\frac{1}{2}(-3\mathcal{P}-4\mathcal{M}-\frac{1}{6}\cK\ell_s\tilde{\rho})\right]\partial_\mu K+i\left[\mathcal{N}-\frac{1}{2}(\ell_s\rho_0+2\mathcal{N}+\frac{3}{2}\mathcal{Q})\right]\partial_\mu K\, .
\label{eq: F-Umu + ansatz}
\end{align} 
\ees

The functions ${\mathcal P}$, ${\mathcal Q}$, $\cM$, $\cN$ are constrained by the extrema conditions resulting from the flux potential \eqref{eq:potentialgeom}, with which they must be compatible. In particular, plugging \eqref{proprhog} into the equations of motion for the axions $\{ \xi^\mu, b^a\}$ one obtains 
\bes
\label{paxionsA}
\begin{equation}
\label{paxioncpxA}
8 \left(\ell_s\rho_0\cM -  {\mathcal P}\cN\right) \partial_\mu K = 0 \, ,
\end{equation}
\begin{equation}
\label{paxionkA}
\left[ 8  {\mathcal P} (\ell_s\rho_0 -  {\mathcal Q}) - \frac{1}{3}\ell_s \tilde{\rho}  \cK \left(-2 {\mathcal Q} + 8 \cN \right)   \right]  \partial_a K   + \left[ \frac{4}{3} \CK\ell_s\tilde{\rho} + 8  {\mathcal P} - 8 \cM \right] \ell_s\rho_{a\mu} u^{\mu} = 0 \, ,
\end{equation}
\ees
which must be satisfied on-shell. Similarly, the equations of motion for the saxions $\{n^\mu, t^a\}$ are

\bes
\label{eq: psaxionsA}
\begin{equation}
\label{eq: psaxioncpxA}
\left(4\ell_s^2\rho_0^2+12\mathcal{P}^3+3\mathcal{Q}^2+8\mathcal{M}^2+8\mathcal{N}^2+\frac{\cK^2}{9}\ell_s^2\tilde{\rho}^2-20\mathcal{QN}-4\mathcal{M}\cK\ell_s\tilde{\rho}\right) \partial_\mu K = 0 \, ,
\end{equation}
\begin{equation}
\label{eq: psaxionkA}
\left[4\ell_s^2\rho_0^2+4\mathcal{P}^2-\mathcal{Q}^2-8\mathcal{QN}+16\mathcal{M}^2-\frac{\cK^2}{9}\ell_s^2\tilde{\rho}^2\right]\partial_a K   + \left[8 \mathcal{Q} -8 {\mathcal N}  \right]\ell_s \rho_{a\mu} u^{\mu} = 0 \, .
\end{equation}
\ees
As it was argued in \cite{Marchesano:2020uqz}, generically \eqref{paxionkA} and \eqref{eq: psaxionkA} imply the constraint 
\be
\rho_{a\mu} u^{\mu} \propto \p_a K\, , 
\label{loophole}
\ee
which was the additional ingredient in the Ansatz used in \cite{Marchesano:2020uqz} for a systematic search of vacua. It turns out that this condition leads to a particular set of SU(3)-structure manifolds,   nearly-K\"ahler manifolds, which is a restricted set of solutions with respect to the class of supersymmetric SU(3)-structure backgrounds obtained in \cite{Lust:2004ig}. In particular, it does not contain the AdS$_4$ vacua constructed in \cite{Caviezel:2008ik,Cribiori:2021djm}, which feature half-flat manifolds.

This suggests that there must be branches of vacua compatible with \eqref{solsfmax} for which \eqref{loophole} is not satisfied, as it is indeed the case. In order to move away from this proportionality relation, one needs to demand that the two pairs of brackets in \eqref{paxionkA} and \eqref{eq: psaxionkA}  vanish simultaneously, which severely constrains the parameters of the Ansatz \eqref{proprhog}. The two possible options are summarised in the two rows of table~\ref{table: vanishing brackets}, where we have defined for clarity the new quantity
\begin{equation}
    \mathcal{S}\equiv 3+4\frac{\mathcal{P}^2}{\mathcal{N}^2}\, .
    \label{eq: param S def}
\end{equation}

\begin{table}[htbp]
\center
\begin{tabular}{|c||c|c|c|c|}
\hline
\diagbox[height=1.25cm,width=4.2cm]{Branch}{Parameters} & $\ell_s \rho_0$ & $\mathcal{Q}$ & $\ell_s\tilde{\rho}\mathcal{K}$ & $\mathcal{M}$ \\ \hhline{|=||=|=|=|=|}
 SUSY&
     $-\frac{3}{2}\mathcal{N}$    &    $\mathcal{N}$            & $-10\mathcal{P}$                          & $-\frac{2}{3}\mathcal{P}$\\   \hline  non-SUSY & $-\frac{\mathcal{N}}{2}\left(1-\frac{12}{\mathcal{S}}\right) $  & $\mathcal{N}$               &          $-6\mathcal{P}\left(1-\frac{4}{\mathcal{S}}\right)$                  &  $\frac{4\mathcal{P}}{\mathcal{S}}$             \\\hline
\end{tabular}
\caption{Parameter configurations of \eqref{proprhog} for which the brackets in \eqref{paxionkA} and \eqref{eq: psaxionkA} vanish independently. At this point there are two branches: one supersymmetric (SUSY) and one non-supersymmetric (non-SUSY).} 
\label{table: vanishing brackets}
\end{table}

By construction, the parameters of table \ref{table: vanishing brackets} satisfy the equations of motion of the Kähler sector. Demanding that this restricted Ansatz also solves the equations of the complex structure sector  further constrains the non-SUSY branch by imposing the vanishing of the Romans mass. As a result, we arrive at the four branches of solutions displayed in table \ref{table: new branches}.

\begin{table}[htbp]
\center
\begin{tabular}{|c||c|c|c|c|c|c|}
\hline
\diagbox[height=1.25cm,width=4.2cm]{Branch}{Parameters} & $\mathcal{P}$ & $\mathcal{S}$ &$\ell_s\rho_0$ & $\mathcal{Q}$ & $\ell_s\tilde{\rho}=m$ & $\mathcal{M}$ \\ \hhline{|=||=|=|=|=|=|=|}
 SUSY&
    Free & \eqref{eq: param S def} &$-\frac{3}{2}\mathcal{N}$    &    $\mathcal{N}$            & $-10\frac{\mathcal{P}}{\cK}$                          & $-\frac{2}{3}\mathcal{P}$\\   \hline \multirow{3}{*}{non-SUSY}& $0$ & $3$ & \multirow{3}{*}{$-\frac{\mathcal{N}}{2}\left(1-\frac{12}{\mathcal{S}}\right)$}   &\multirow{3}{*}{$\mathcal{N}$}     &    \multirow{3}{*}{$0$  }   &  \multirow{3}{*}{$\frac{4\mathcal{P}}{\mathcal S}$}             \\\cline{2-3}  & $+\frac{\mathcal{N}}{2}$ & 4 &    &      &              & \\\cline{2-3} & $-\frac{\mathcal{N}}{2}$ &  4 &   &              &     &   \\\hline
\end{tabular}
\caption{Complete set of branches of solutions to the equation of motion under the Ansatz \eqref{proprhog} such that the brackets in \eqref{paxionkA} and \eqref{eq: psaxionkA} vanish simultaneously. The SUSY branch has two independent parameters $\mathcal{P}$ and $\mathcal{N}$ while the non-SUSY branches only have one, i.e. $\mathcal{N}$.} 
\label{table: new branches}
\end{table}

By evaluating the above solutions in the F-terms equations \eqref{eq: F-terms + ansatz} one can check that the first row of table \ref{table: new branches} corresponds to a supersymmetric branch of vacua. Actually, these results generalise the SUSY branch found in \cite[table 2]{Marchesano:2020uqz} beyond the case \eqref{loophole}. In addition, we also find three new non-SUSY families of solutions. From the 10d perspective, addressed in section \ref{sec:10d}, these four branches describe half-flat manifolds, which contrasts with the nearly-Kähler geometry arising when \eqref{loophole} is satisfied.  

Motivated by the T-duality considerations of section~\ref{s:Tdual}, from now on we focus on the solutions with vanishing Romans mass, which leads us take $ {\mathcal P}=\cM = 0$ for the SUSY branch. We also focus on the non-SUSY branch that features ${\mathcal P}=\cM = 0$ since it shares a lot of similarities with its SUSY partner. From  table \ref{table: new branches} we thus read the condition
\be
\tilde{\rho} = {\mathcal P} = \cM = 0\, ,
\label{Ansatz1}
\ee
and that the SUSY sub-branch and its non-SUSY counterpart are described by:
\bes
\label{branches}
\begin{align}
    \textrm{SUSY:}  & \qquad \quad \mathcal{Q}=\mathcal{N}\,, \qquad \ell_s\rho_0=-\frac{3}{2}\mathcal{N}\,,\\
    \textrm{non-SUSY:}  & \qquad \quad   \mathcal{Q}=\mathcal{N}\,, \qquad \ell_s\rho_0=\frac{3}{2}\mathcal{N}\,. 
    \label{nosusybranch}
\end{align}
\ees

Both branches display a negative vacuum energy given by\footnote{ Note that here we write the vacuum energy like $V$ which differs from $V_F$ of expression \eqref{eq:potentialgeom} in that it is now in Einstein frame and, in addition, it is expressed in four-dimensional Planck units.}
\begin{equation}
    V|_{\rm vac}=-12e^K\mathcal{Q}^2\, , 
    \label{eq: scalar potential vacuum new ansatz}
\end{equation}
which is independent of the choice of sign in \eqref{branches}.  This sign choice is however physically relevant, since it distinguishes between supersymmetric and non-supersymmetric vacua, as we argued before. In the non-supersymmetric case, the F-term structure corresponds to \eqref{solsfmax} with\footnote{ Again here we abuse notations and $W$ is now in Einstein frame and expressed in four-dimensional Planck units.}
\be
\ell_s \lambda_K = \ell_s \lambda_Q =  3\mathcal{Q}\, , \qquad  e^K|W|^2_{\rm vac} =  25 e^K \mathcal{Q}^2\, ,
\label{Ftermnosusy}
\ee
which indeed reproduces \eqref{eq: scalar potential vacuum new ansatz}. In other words, we find two branches of vacua related by a sign flip, which preserves the vacuum energy but breaks supersymmetry. This is reminiscent of the structure found in DGKT-CFI vacua (see also \cite{Marchesano:2019hfb}) where the sign flip is implemented in the (smeared) RR four-form flux background $G_4$. This is a first hint that these new branches may contain the T-duals of DGKT-CFI vacua discussed in the section \ref{s:Tdual}.  Further evidence can be gathered by providing a 10d interpretation of these 4d solutions, as we now proceed to do.

\subsection{10d interpretation}
\label{sec:10d}

As emphasised in \cite{Bielleman:2015ina,Carta:2016ynn,Herraez:2018vae}, the flux-axion polynomials $\{\rho_0, \rho_a, \tilde{\rho}^a, \tilde{\rho}\}$ represent the type IIA gauge invariant RR fluxes $G_{2p}$ threading the compactification manifold $X_6$ with $p=3,2,1,0$ respectively,  as opposed to the flux quanta $\{e_0,e_a, m^a, m\}$.  Additionally, one may connect with the results of  \cite{Lust:2004ig,Caviezel:2008ik,Cribiori:2021djm} by interpreting the metric fluxes $\rho_{a\mu}$ in terms of the intrinsic torsion classes of the SU(3)-structure manifold $X_6$. In order to proceed, we make the assumption that the presence of fluxes causes a small deformation of the internal geometry such that the underlying Calabi--Yau structure can still be used as a reference, as in \cite{Gurrieri:2002wz,DAuria:2004kwe,Tomasiello:2005bp,Grana:2005ny,Kashani-Poor:2006ofe}. In particular, we  have a well-defined Kähler potential that satisfies the standard relations 
\begin{equation}
    \cK\partial_\mu K \beta^\mu=-6 \, e^{\phi}\re\Omega\,,\qquad \cK\partial_a K \tilde{\omega}^a=3J\wedge J\,.
\end{equation}
Using them, the definition \eqref{deffamu} and the Ansatz \eqref{proprhog} and \eqref{branches}, we have that
\begin{align}
    \ell_s t^a\rho_{a\mu}&=\mathcal{Q}\partial_\mu K &\Longrightarrow & & \dd J= t^a \dd\omega_a&= \frac{6\mathcal{Q}}{\cK\ell_s}\re\Omega\\
     \ell_s \rho_{a\mu}u^\mu&=\mathcal{Q}\partial_a K-\ell_s\cK_{ab}\tilde{\rho}^b&\Longrightarrow & & \dd\im\Omega =e^{\phi}u^\mu \dd\alpha_\mu &=\left(-3\frac{\mathcal{Q}}{\cK\ell_s}J^2-J\wedge G_2\right) e^{\phi}\, .
\end{align}

Then, in order to derive the torsion classes we only need to decompose the flux $G_2$ in its primitive $G_2^{\rm P}$ and non-primitive $G_2^{\rm NP}$ components 
\begin{equation}
    G_2=G_2^{\rm P}+G_2^{\rm NP}\,,
\end{equation}
with $G_2^{\rm P}\wedge J^2=0$. The Ansatz \eqref{proprhog} tells us that $\ell_s\cK_a\tilde{\rho}^a=\mathcal{Q}$, which implies 
\begin{equation}
    G_2^{\rm NP}=\frac{\mathcal{Q}}{\cK\ell_s}J\,,\qquad 
     G_2^{\rm P}=-\frac{4\mathcal{Q}}{\cK\ell_s}J-\cK^{ab}\rho_{b\mu}u^\mu \omega_a\,.
\end{equation}

From the above discussion we deduce that when the localised sources are smeared (so that the Bianchi identities amount to the tadpole condition \eqref{BIG2}, already taken into account by our analysis), the branches in table \ref{table: new branches} correspond to the following 10-dimensional flux configurations and SU(3)-structure relations
\begin{equation}
  \begin{gathered}
     G_0 = -\frac{10\mathcal{P}}{\cK\ell_s}\,, \qquad G_4 = -\frac{3\mathcal{P}}{\cK\ell_s}J^2\,,\qquad  H = -\frac{4\mathcal{P}}{\cK\ell_s}e^\phi \re \Omega\,,\\
     G_6=-\rho_0\Phi_6\,,\qquad  G_2=\frac{\mathcal{Q}}{\cK\ell_s}J+ G_2^{\rm P}\,,
\end{gathered}  
\end{equation}
\begin{equation}
    \dd J = \frac{6\mathcal{Q}}{\cK\ell_s}e^\phi \re \Omega\,,\qquad 
    \dd\im\Omega = \left(-\frac{4\mathcal{Q}}{\cK\ell_s}J^2-J\wedge G_2^{\rm P} \right) e^\phi \, , \label{eq: external derivatives of J and Omega} 
\end{equation}
where $\Phi_6$ is the normalised volume 6-form and $\rho_0$ takes the value $-\frac{3}{2} \mathcal{Q}$ in the SUSY branch and $-\oh\mathcal{Q} (1-12/\mathcal{S})$ in the non-SUSY branches. The two relations in \eqref{eq: external derivatives of J and Omega} translate into the following $SU(3)$ torsion classes
\begin{equation}
    \mathcal{W}_1=-i\frac{4\mathcal{Q}}{\cK\ell_s} e^\phi \,,\qquad \mathcal{W}_2=-iG_2^{\rm P}e^{\phi} \,,\qquad \mathcal{W}_3=\mathcal{W}_4=\mathcal{W}_5=0\,.
\end{equation}

Therefore, in terms of an internal SU(3)-structure manifold, our vacua correspond to a subfamily of half-flat compactifications with $\mathcal{W}_3=0$, like the ones considered in \cite{Lust:2004ig,Koerber:2008rx, Caviezel:2008ik,Cribiori:2021djm}. This complements the picture found in  \cite{Marchesano:2020uqz}, where it was observed that flux configurations satisfying \eqref{loophole} describe nearly-Kähler compactifications. From this framework, one can interpret the vacuum constraint \eqref{Ansatz1} as the absence of certain internal fluxes:
\be
G_0 = G_4 = H = 0\, ,
\label{noflux}
\ee
while $G_2$ and $G_6$ are present in the compactification, together with metric fluxes. This is indeed what we expect to find after applying two T-dualities to a DGKT-CFI vacuum, at least in the approximation of smeared localised sources. The absence of Romans mass and $H$-flux was already pointed out in section \ref{s:Tdual}, while the absence of  $G_4$ in the smearing approximation and the presence of $G_2$ and $G_6$ can be motivated by looking at the solutions found in \cite{Caviezel:2008ik,Cribiori:2021djm}. 

 We close this section by recalling that the current discussion does not mean that the internal metric of $X_6$ corresponds to an SU(3)-structure. As in the 10d uplift of the 4d supersymmetric Calabi--Yau vacua \cite{DeWolfe:2005uu}, recently analysed in \cite{Junghans:2020acz,Marchesano:2020qvg}, one expects that upon localisation of sources, the 10d background displays an 
 SU(3)$\times$SU(3)-structure that is simplified to an SU(3)-structure in the smearing approximation. An example that motivates this picture in the context of  half-flat manifolds are the twisted-tori solutions with localised sources obtained in \cite{Cribiori:2021djm}.


\section{Moduli stabilisation and scaling symmetries}
\label{s:moduli}

In this section we analyse the vacua equations for the two branches obtained in \eqref{branches}, in order to work out explicit expressions for the fixed moduli vevs  in terms of flux quanta. We show how, in simple cases, one recovers scaling symmetries of the form \eqref{newsc}, that lead to an infinite family of vacua with the scale-separation relations \eqref{scalingsep}. Two key assumptions to find such scaling symmetries are a  metric-flux matrix of rank one and a compactification manifold of the form $X_6 = (T^2  \tilde{\times} X_4)/\Gamma$, where $X_4 = T^4$ or $K3$, $\Gamma$ is an orbifold group and $\tilde{\times}$ indicates that metric fluxes fibre $T^2$ over $X_4$. These examples are directly related to the ones considered in \cite{Caviezel:2008ik,Cribiori:2021djm} for $X_4 =T^4$ and in \cite{Lust:2004ig} for $X_4 = K3$. In section \ref{s:elliptic} we  extend our 4d analysis to manifolds that are more general elliptic fibrations, and show that for them the scaling symmetry is no longer exact. Nonetheless, we will argue that such geometries can also lead to an infinite family of vacua with a parametric scale separation similar to the cases discussed in here.

\subsection{A rank-one metric flux Ansatz}

Let us start by rewriting the on-shell relations \eqref{Ansatz1} and \eqref{branches} as:
\begin{subequations}
\begin{align}
\label{vanirho}
\tilde{\rho}=\rho_\mu=\rho_a=&\, 0\, ,\\
    \ell_s (\mathcal{K}_{ab}\tilde{\rho}^b+\rho_{a\mu}u^\mu) =&\, \mathcal{Q}\partial_a K\, , \label{eq: pa}\\
    \ell_s \rho_{a\mu}t^a=&\, \mathcal{Q}\partial_\mu K\, ,\label{eq: pmu}\\
    \ell_s \rho_0=&\pm\frac{3}{2}\mathcal{Q}\, ,
\end{align}
\label{eq: rhos in the new ansatz}
\end{subequations}
where a priori $\mathcal{Q}$ can be function of the moduli. The negative sign corresponds to the supersymmetric branch of vacua \eqref{branches}, and the positive to the non-supersymmetric one. If these vacua are obtained by applying two T-dualities to a DGKT-CFI vacuum, one expects a metric flux matrix $f_{a\mu}$ of rank one, as observed in \eqref{newWNS}. Let us then take this as a working assumption,\footnote{Interestingly, finding flux scaling symmetries like the ones in the next subsection turns out to be quite involved for an $f_{a\mu}$ of higher rank, due to the additional tadpole constraints that appear.} and write
\be
f_{a \mu} \equiv \sigma_a \sigma_\mu\, ,
\label{famufact}
\ee
with $\sigma_a, \sigma_\mu \in \IZ$. Then, in order to satisfy the vacuum condition $\rho_\mu=0$ from \eqref{eq:rhomu}, we necessarily need to impose the following condition on the fluxes
\be
h_\mu =  \sigma \sigma_\mu \quad \implies \quad \sigma_a b^a|_{\rm vac} = - \sigma\, , 
\label{hprop}
\ee
with $\sigma \in \IZ$, which partially fixes some axion directions. To proceed we consider the matrix $J_{ab} \equiv \cK_{abc}m^c$ and impose a second working assumption:
\be
\sigma_a = J_{ab} \chi^b\, , \qquad \text{for some} \ \chi^b \in \IR\, .
\label{sigcond}
\ee 
Namely, that the vector $\sigma_a$ is in the image of the symmetric matrix $J$. This assumption is automatically satisfied whenever $J$ is invertible, which will be the case in all of our examples. 

Under the assumption \eqref{sigcond} and the condition $\tilde{\rho} = 0$, the vacuum equation $\rho_a= 0$ in \eqref{eq:rho_ag} can only be satisfied if also $e_a = J_{ab} \zeta^b$ for some $\zeta^b\in \IR$. It follows that $\rho_a= 0$ translates into 
\be
b^a|_{\rm vac}  = - \zeta^a - \chi^a  \sigma_\mu \xi^\mu|_{\rm vac}\, ,
\label{bavac}
\ee
which implies that those K\"ahler axions in the kernel of $J$ are left unfixed. Note also that only the complex structure axions that enter the linear combination $\sigma_\mu \xi^\mu$ are fixed by the vacuum equations, although these axions are rendered massive via St\"uckelberg couplings involving space-time filling D6-branes \cite{Camara:2005dc}. Using both \eqref{hprop} and \eqref{bavac}, we find that the vev for such a combination is
\be
\label{ximuvac}
  \sigma_\mu \xi^\mu|_{\rm vac} =   \frac{\sigma -\sigma_a \zeta^a }{ \sigma_a\chi^a}  \, \stackrel{\exists J^{-1}}{=} \, \frac{\sigma -J^{ab}\sigma_a e_b }{ J^{ab}\sigma_a\sigma_b} \, ,
\ee
where in the second equation we have assumed that $J$ is invertible, with inverse $J^{ab}$. Plugging this expression into the definition of $\rho_0$ in \eqref{eq:rho0g} and using \eqref{vanirho} we finally find
\be
\ell_s\rho_0|_{\rm vac} = e_0 - \oh J^{ab} e_ae_b  + \oh  \frac{(\sigma -J^{ab}\sigma_a e_b)^2 }{ J^{ab}\sigma_a\sigma_b}\, ,
\label{rho0vac}
\ee
and the corresponding expression in terms of $(\chi^a, \zeta^b)$ when $J$ is not invertible. That is, we have found explicit formulas for all the stabilised axion vevs and for ${\mathcal Q}$ in terms of fluxes quanta. Recall that this last value essentially fixes the vacuum energy via \eqref{eq: scalar potential vacuum new ansatz}. In simple examples, like for instance the backgrounds considered in  \cite{Cribiori:2021djm}, we have that $e_a = h_\mu =0$, and then ${\mathcal Q}^2  = \frac{4}{9} e_0^2$. 

Let us now consider the vacuum equations involving the saxions, namely \eqref{eq: pa} and \eqref{eq: pmu}. We first have that \eqref{eq: pmu} implies the on-shell relations
\be
 {\mathcal Q} \p_\mu K|_{\rm vac}  = \sigma_\mu \sigma_a t^a |_{\rm vac}   \quad \implies \quad \sigma_\mu u^\mu|_{\rm vac} =  -4 {\mathcal Q} (\sigma_a t^a)^{-1}_{\rm vac} \, ,
 \label{sax1}
\ee
which when plugged back into \eqref{eq: pa} leads to
\be
J_{ab} t^b|_{\rm vac} =  {\mathcal Q} \left[ 4 \frac{\sigma_a}{\sigma_a t^a} -3 \frac{\cK_a}{\CK}\right]_{\rm vac} \quad \implies \quad J_{ab} t^a t^b|_{\rm vac} =  {\mathcal Q}\, .
\label{sax2}
\ee
The first of these last relations represents a non-linear equation on the K\"ahler saxions which, when solved, can be plugged back into \eqref{sax1} to obtain the vev of the complex structure saxions.

\subsection{Scaling symmetries in twisted factorised geometries}
\label{sec:factorized}

Solving the saxion vacuum equations \eqref{sax1} and \eqref{sax2} explicitly is in general non-trivial, and it is also not clear how the dependence of the saxion vevs on the fluxes can lead to a scaling symmetry of the form \eqref{newsc} that generates an infinite family of solutions. At this stage the only general information that we have is that the fluxes with a non-trivial contribution to the Bianchi identity of $G_2$   cannot scale, which in this case corresponds to the product $m^af_{a\mu} = m^a\sigma_a \sigma_\mu$. This means that one can  scale arbitrarily ${\cal Q}$ and the flux quanta $m^a$ that do not enter in $m^a\sigma_a$. 

The simplest way to generate an infinite family of vacua from such a flux scaling is whenever there is a scaling symmetry in the vacuum equations. Looking at \eqref{sax2} this looks difficult to implement unless the triple intersection numbers of $X_6$ factorise as
\be
\cK_{abc} = \sigma_{(a} \eta_{bc)} \, ,
\label{Kfact}
\ee
with $\eta_{ab}=\eta_{ba}$ and $\sum_a\sigma_a\eta_{ab}=0$. Note that we do not include any normalisation factor in our definition of the symmetrisation of the indices. This structure simplifies \eqref{sax2} to
\be
m^b \eta_{bc} t^c|_{\rm vac} = \frac{3\cal Q}{t^L|_{\rm vac}} \qquad \text{and} \qquad \eta_{ab}t^at^b|_{\rm vac}  =  - \frac{5\cal Q}{m^a\sigma_a}\, ,
\label{simpsax2}
\ee
where we have dubbed $t^L \equiv t^a\sigma_a$. We can then implement the scalings 
\be
{\cal Q} \sim n^{2r}\, , \qquad m^b\eta_{bc} \sim n^{r-s}\, , \qquad m^a\sigma_a \sim \text{const.}
\ee
with $r \geq s \geq 0$, that lead to\footnote{As mentioned earlier, when $e_a=h_\mu=0$, ${\cal Q}$ is proportional to $e_0$ and the exact symmetry is obtained simply by scaling $e_0\sim n^{2r}$. In general, ${\cal Q}$ takes the form \eqref{rho0vac}. It is however still possible to generate the said scaling if the remaining flux quanta scale like $e^a\sigma_a\sim n^{2r-s}$, $e^a\eta_{ab}\sim n^r$ and $h_\mu\sim n^s$.}
\be
t^b \eta_{bc} \sim n^r\, ,\qquad t^L \sim n^s \, , \qquad u^\mu \sim n^{2r-s}\, , 
\label{temp}
\ee
and which reproduce \eqref{newsc} and \eqref{newsc2} except for a sign flip on $s$ expected from T-duality. With these scalings one can check that
\be
 R_{\rm AdS}  M_{\rm KK} \sim n^{ \oh (r-s) } \, ,
 \label{scalingsep2}
\ee
and so scale separation is obtained whenever $r > s \geq 0$. 

To illustrate this picture it is useful to consider a particular geometry for $X_6$. Clearly, the structure \eqref{Kfact} for the triple intersection numbers occurs for factorised geometries like $T^2 \times X_4$. However, these may not be acceptable geometries to solve the Bianchi identities. Indeed, recall that in the smeared approximation the exterior derivative for the RR two-form flux threading $X_6$ is given by \eqref{BIG2} with $\rho_\mu = 0$. Due to \eqref{sax1} we have that $m^a f_{a\mu} \b^\mu \propto \re \Omega$, and so one needs an O6-plane content such that  $\delta_{\rm O6} \propto  \re \Omega$. This cannot be achieved in factorised geometries, unless the compactification is modded out by an appropriate orbifold group $\Gamma$, as done in \cite{Caviezel:2008ik} for the case of nilmanfolds. Therefore in the following we will focus on geometries of the form $X_6 = (T^2 \tilde{\times} X_4)/\Gamma$, where $X_4 = T^2 \times T^2$ or $K3$, the symbol $\tilde{\times}$ means that we are twisting the product by introducing a rank-one metric flux $f_{a\mu}$ that fibres the $T^2$ over $X_4$, and $\Gamma$ is a orbifold quotient such that $\delta_{\rm O6} \propto  \re \Omega$. Finally, in order to recover a structure of the form \eqref{Kfact}, we will focus our attention on the orbifold untwisted sector of the theory. 
Notice in this setting there is no general scheme to stabilise the orbifold twisted sector away from the orbifold point, and as a result the final compactification will typically display orbifold singularities.

For both cases of $X_4$ we can split the K\"ahler index as $a =\{ L, A\}$, where $L$ refers to $T^2$ and $A$ runs over the K\"ahler moduli of $X_4$. We have that
\be
\label{tripfac}
\CK_{LAB} = \kappa \eta_{AB}\, , \qquad \CK_{ABC} =\CK_{LLA} = \CK_{LLL} = 0\, , 
\ee
where $\kappa \in \mathbb{N}$ and $\eta_{AB}$ is the intersection matrix of $X_4$.  This implies
\be
\CK = 3 t^L \kappa \eta_{AB}t^At^B = 3t^L \kappa\hat\eta\, .
\ee
where we have defined $\hat\eta \equiv \eta_{AB}t^At^B$. The structure \eqref{Kfact} is recovered if we take $\sigma_a = \sigma_L  \delta_{aL}$ with $\sigma_L=\kappa$. Then, eq.~\eqref{sax1} amounts to 
\be
\sigma_\mu u^\mu|_{\rm vac} = -\frac{4{\cal Q}}{\sigma_L t^L}\, .
\ee
In addition we have that
\be
J_{ab} \equiv \CK_{abc}m^c =\sigma_L 
\begin{pmatrix}
0 & \eta_{AB}m^B \\  \eta_{AB}m^B & m^L  \eta_{AB}
\end{pmatrix}\, ,
\label{Jtriv}
\ee
\be
J^{ab}  = \frac{1}{\sigma_L M}
\begin{pmatrix}
- m^L	& m^A \\ m^A & - \frac{m^Am^B}{m^L} + \frac{\eta^{AB}}{m^L} M 
\end{pmatrix}\, ,
\label{Jinvtriv}
\ee
where we have defined $M \equiv \eta_{AB}m^Am^B$. Since $J_{ab}$ is invertible we can rewrite \eqref{sax2} as the more explicit expressions
\bes
\label{teqs}
\begin{align}
\label{tL1}
t^L & =  -\frac{{\cal Q}}{\sigma_L M} \left(3\frac{ m^L}{t^L} +2\frac{m^A\eta_A}{\hat\eta} \right)  \, ,\\
t^A & = \frac{{\cal Q}}{\sigma_L M} \left(3\frac{m^A}{t^L}   + 2 m^A  \frac{m^B\eta_B}{m^L\hat\eta} - 2 \frac{M}{m^L} \frac{t^A}{\hat\eta}\right) \, ,
\label{tA1}
\end{align}
\ees
where $\eta_A \equiv \eta_{AB} t^B$. This system is solved by
\be
\label{vevsfact}
t^A = - \frac{5}{3} \frac{m^A}{m^L} t^L\quad \text{with} \quad t^L = \sqrt{-\frac{9{\cal Q} m^L}{5\sigma_L M}} \, ,
\ee
which is in agreement with \eqref{simpsax2} and realises the expected scalings \eqref{temp} via the following rescaling of flux quanta
\be
{\cal Q} \sim n^{2r}\, , \qquad m^A \sim n^{r-s}\, , \qquad m^L \sim \text{const.}
\label{eq: scale separated flux scalings}
\ee
This also reproduces the results presented in \cite{Cribiori:2021djm}, for the choice of scaling parameters $a = 2r$ and $b=c=r-s$ defined therein.
In order to maintain the consistency of the scaling symmetry across the remaining equations of motion \eqref{bavac}, \eqref{ximuvac} and \eqref{rho0vac} we also take
\begin{equation}
\label{scalings_Q}
    e_0\sim n^{2r}\,,\quad e_A\sim n^r\,,\quad e_L\sim n^{2r-s}\,,\quad h_\mu \sim n^s\,.
\end{equation}
Given that $e^{-K_Q}$ is a homogeneous function of degree four and using $u^\mu \sim n^{2r-s}$ we deduce 
\begin{equation}
    e^{-K_Q}=e^{-4\phi_4}\implies u^\mu\sim e^{-\phi_4}  \implies e^{\phi_4}\sim n^{-2r+s}\,,
\end{equation}
which in turn leads to the following scaling for the 10d string coupling
\begin{equation}
\label{gs}
    g_s\sim e^{\phi_4}\sqrt{{\rm Vol}_{X_6}} \sim n^{-2r+s} \cdot n^{r+\frac{s}{2}}=n^{-r+\frac{3}{2} s}\,.
\end{equation}
Consequently in order to stay in the weak coupling limit one further needs to require $r>\frac{3}{2} s$, in agreement with the results of \cite{Cribiori:2021djm}.

The Kaluza-Klein scale is approximated by the largest compact direction, which is contained in $X_4$. Therefore the corresponding scaling reads
\begin{equation}
    L_{\rm KK} \sim \sqrt{t^A}\sim n^{r/2}\,, \qquad M_{\rm KK}= (\ell_s L_{\rm KK})^{-1} \implies \frac{M_{\rm KK}^2}{M_{\rm P}^2} \sim \frac{g_s^2}{{\rm Vol}_{X_6} t^A}\sim n^{-5r+2s}\,,
\end{equation}
where in the last step we have expressed the result in units of the Planck mass $M_{\rm P}$. The AdS scale can be obtained from the scalar potential \eqref{eq: scalar potential vacuum new ansatz}
\begin{equation}
\Lambda=(R_{\rm AdS}M_{\rm P})^{-2}\sim e^K \mathcal{Q}^2\sim \frac{\mathcal{Q}^2}{(u^\mu)^4t^A t^B t^L}\sim n^{-6r+3s}.
\label{Hubblescale1}
\end{equation}
Therefore we conclude

\begin{equation}
    R_{\rm AdS}  M_{\rm KK}\sim n^{\oh(r-s)}\,,
\end{equation}

which means that this new branch of solutions provides scale separated vacua whenever $r>s\geq 0$ in the choice of flux rescaling \eqref{eq: scale separated flux scalings}, as predicted in \eqref{scalingsep2}. Finally, let us verify that higher order flux corrections are under control in the regime in which scale separation arises. Following \cite{Caviezel:2008ik}, we compute the norm $|g_sG_p|^2$ of the flux density of the fields that are present in our setup, i.e. $G_2$ and $G_6$. Applying the scalings described above we have
\begin{align}
    |G_2|^2g_s^2\sim \cK_{ab}m^am^b \frac{g_s^2}{\rm Vol_6} \sim n^{2r-s}\cdot n^{-2r+3s}\cdot n^{-2r-s} \sim n^{s-2r}\,,\\
   |G_6|^2g_s^2\sim \frac{e_0^2}{\sqrt{|g|}} \cdot \frac{g_s^2}{\rm Vol_6}\sim n^{2r-s}\cdot n^{-2r+3s}\cdot n^{-2r-s}\sim n^{s-2r}\,,
\end{align}
and therefore we conclude that this corrections are suppressed in the regime of large $n$ whenever $2r>s\geq0$, which is compatible the scenario required to achieve scale separation. 


\section{Elliptic fibrations}
\label{s:elliptic}

We now generalise the previous results to the more involved case of smooth elliptically fibered geometries $X_6$. The possibility to extend the results of \cite{Cribiori:2021djm} and obtain similar scalings in this context has already been anticipated at the end of section \ref{s:Tdual}, and the general idea goes as follows. Let us consider the DGKT-like set of vacua based on a smooth elliptically fibered Calabi--Yau. One can then single out a four-form flux quantum dual to the class of the elliptic fibre and scale it like $\hat{e}_1$ in \eqref{newsc}, while all the remaining four-form fluxes scale like $\hat{e}_i$. As in the toroidal case, for negative values of $s$ one expects to go to a small fibre area limit. Therefore, by performing a double T-duality along the fibre, or in other words a Fourier-Mukai transformation, one should be taken to a large-volume type IIA compactification on a different elliptic fibration $X_6$. From the Fourier-Mukai action on D-brane charges (see e.g. \cite{Corvilain:2018lgw}) one deduces that at the 4d EFT level this duality must be implemented by transformations of the form \eqref{Tinv} and \eqref{Ktrans}, and that most of the features of the toroidal setting should also apply to this case. In particular, the transformed background contains no Romans mass, and the $H$-flux must be traded by metric fluxes. The corresponding flux vacua should then lie within the branches \eqref{Ansatz1} and \eqref{branches}.  

Nevertheless, in this setup the analogy with toroidal compactification is not exact, due to the more involved structure of the triple intersection numbers, that no longer factorises. It is therefore a natural question whether or not one may achieve an infinite family of vacua with parametric scale separation via similar flux scalings. To address this question we will first describe an exact scaling symmetry of the vacuum equations to see how we one can reproduce the scalings \eqref{newsc} and \eqref{newsc2} in this more involved setting. It turns out that such a scaling symmetry is however only formal, with little hope to be implemented in practice, since one would need to scale topological data of $X_6$ and as a result an infinite family of vacua cannot be generated. In contrast, one can generate such an infinite family by implementing an approximate scaling symmetry that reduces to \eqref{newsc} and \eqref{newsc2} at leading order in a perturbative expansion. This expansion leads to a perturbative approach to solve the saxionic vacuum equations around the saxionic vevs of the trivially fibered case analysed in the previous section, which is a more and more accurate approximation as we move along the infinite family of vacua.  We will eventually refer to numerical results obtained in specific examples and displayed in Appendix ~\ref{ap:numerical}, to confirm the robustness of the perturbative analysis and the existence of this new infinite families of scale-separated vacua on smooth elliptically fibrations.

\subsection{An exact scaling symmetry}

Let us assume that one can build a DGKT-like family of vacua on a smooth elliptically fibered Calabi--Yau $W_6$ with base $X_4$. As in section \ref{sec:factorized}, we split the K\"ahler index as $a =\{ L, A\}$, where $L$ refers to the elliptic fibre and $A$ runs over the K\"ahler moduli of the base $X_4$. Expanding the first Chern class of the base like $c_1(X_4)\equiv c^A\omega_A$, the triple intersection numbers read
\begin{equation}
\label{kappafib}
\cK_{LAB}=\eta_{AB}\, ,\quad \cK_{LLA}=\eta_{AB}c^B\, ,\quad \cK_{LLL}=\eta_{AB}c^Ac^B\, ,\quad \cK_{ABC}=0\, .
\end{equation}
which implies 
\begin{equation}
\cK=\eta_{AB}t^L\left(c^Ac^B(t^L)^2+3c^At^Bt^L+3t^At^B\right)\, .
\end{equation}

We now perform two T-dualities along the fibre or, equivalently, a Fourier-Mukai transform. Following the reasoning of section \ref{s:general} and the results of \cite{Fidanza:2003zi,Tomasiello:2005bp}, the former $H$-flux $H \propto \re \Omega$ should be translated into a more involved elliptic fibration that leads to an SU(3)-structure manifold $X_6$, such that it has same structure of triple intersection numbers as $W_6$, but now with non-trivial intrinsic torsion classes. This last feature should be captured by a rank-one metric flux matrix of the form \eqref{famufact}, and more precisely such that  $\sigma_a= \sigma_L \delta_{aL}$. It is in this manifold $X_6$ where we are going to look for solutions to the 4d vacua equations \eqref{Ansatz1} and \eqref{branches}. 

The discussion parallels the one in section \ref{sec:factorized}, except that now we have the more involved triple intersection numbers \eqref{kappafib}. This implies that the matrix $J_{ab}$ and its inverse read
\begin{equation}
J_{ab}\equiv\cK_{abc}m^c=\begin{pmatrix}
\eta_{AB}\left(c^Ac^Bm^L+c^Am^B\right) & \eta_{AB}(m^B+c^Bm^L)\\
\eta_{AB}(m^B+c^Bm^L) & \eta_{AB}m^L
\end{pmatrix},
\end{equation}
\begin{equation}
J^{ab}=\frac{1}{N}\begin{pmatrix}
-m^L & m^A+c^Am^L\\
 m^A+c^Am^L & -\frac{m^Am^B}{m^L}+\frac{\eta^{AB}}{m^L}N-c^{(A}m^{B)}+c^Ac^Bm^L
\end{pmatrix},
\end{equation}
with $N\equiv M+\eta_{AB}c^Am^Bm^L$. Notice that we recover \eqref{Jtriv} and \eqref{Jinvtriv} with $\kappa=1$ when the Chern class of the base vanishes. Moreover, with this choice of metric flux  eq.~\eqref{sax1} becomes 
\be
 {\mathcal Q} \p_\mu K|_{\rm vac}  = \sigma_\mu \sigma_L t^L |_{\rm vac}   \quad \implies \quad \sigma_L\sigma_\mu u^\mu|_{\rm vac} =  -\frac{4 {\mathcal Q}}{t^L |_{\rm vac}} \, ,
 \label{sax1fib}
\ee
while eq.~\eqref{sax2} multiplied by $J^{-1}$ yields
\begin{equation}
\cK t^a=-3\mathcal{Q}J^{ab}\cK_b+\frac{4\mathcal{Q}\cK}{t^L}J^{aL}\, .
\end{equation}
Splitting this relation for the index $L$ and indices $A$ we get
\bes
\begin{align}
\cK t^L&=-\frac{\mathcal{Q}}{N}\left\{m_L\left[9\hat\eta+4(c^t\eta c))(t^L)^2+12(c^t\eta t)t^L\right]+3(m^t\eta c)(t^L)^2+6(m^t\eta t)t^L\right\}\, ,\\
\cK t^A&=\frac{\mathcal{Q}}{N}\left\{-\frac{3Mt^L}{m^L}\left(2t^A+c^At^L\right)+3(c^t\eta m)t^L\left(2t^A-\frac{m^A}{m^L}t^L\right)\right.\\
&\hspace{1.12cm}\left.+(m^A+c^Am^L)\left[9\hat\eta+6\frac{m^t\eta t}{m^L}t^L+4(c^t\eta c)(t^L)^2+12(c^t\eta t)t^L\right]\right\}\nonumber\, ,
\end{align}
\label{saxfib}
\ees
where $x^t\eta y\equiv \eta_{AB}x^Ay^B$. In particular, $M=m^t\eta m$ and $\hat\eta=t^t\eta t$.

From the system of equations above, we can derive an exact scaling symmetry of the form
\begin{equation}
\label{scalingsfib}
m^A\sim n^{r-s}\, ,\qquad m^L\sim \text{const.}\, ,\qquad \mathcal{Q}\sim n^{2r}\, ,\qquad c^A\sim n^{r-s}\ ,
\end{equation}
which gives exactly the desired scalings \eqref{newsc} and \eqref{newsc2} for the saxions to obtain scale separation. This symmetry is however only formal since it involves scaling the first Chern class $c^A$ of the base of the elliptic fibration, which cannot be done in general. Nevertheless, the observation that the Chern class needs to scale with a positive power of $n$ (recall that we take $r>s$) in order to have an exact symmetry hints at the fact that the terms breaking the symmetry when we keep $X_4$ fixed are subleading in the limit of large $n$. This is supported by the fact that each Chern class term in \eqref{saxfib} is multiplied by additional factors $\{m^L, t^L\}$ in detriment of $\{m^A,t^A\}$. Therefore, if the scalings of section \ref{sec:factorized} were to be taken, the terms associated with the first Chern class of the base would grow slower than the ones associated to a factorised geometry with a damping power $n^{s-r}$. In the following, we will consider this scenario in more detail and perform a perturbative expansion around the factorised limit.

\subsection{An approximate scaling symmetry through a perturbative expansion}
\label{pert}

Let us now implement the scaling of the   flux quanta (that are not involved in the tadpole) while keeping the fibration base $X_4$ fixed. In this case the scaling symmetry \eqref{scalingsfib} is broken, and so it is not clear that one can  generate an infinite family of scale-separated vacua. To argue that this is  the case, we resort to a perturbative expansion of the 4d equations of motion, related to an approximate scaling symmetry. Indeed, recall that the scaling $m^A\sim n^{r-s}$ leads to an exact symmetry either when $c^A$ vanishes or it is rescaled as in \eqref{scalingsfib}. When it is non-vanishing and fixed, one can define the expansion parameters $\epsilon^A$ to be
\begin{equation}
\epsilon^A \equiv \frac{c^A}{m^A}\sim n^{s-r}\, ,
\end{equation}
which are all equally suppressed for large values of $n$ and $r > s \geq 0$.  

One can now solve the 4d vacua equations order by order in $\eps$. In particular, inserting the vev solutions \eqref{tL1} and \eqref{tA1} of the factorised case into the relations \eqref{saxfib} turns out to cancel the leading order terms. We therefore write the perturbative expansion around this solution as
\begin{equation}
t^L\equiv t^L_{(0)}\left(1+\Delta^L+\mathcal{O}(\epsilon^2)\right)\, ,\qquad
t^A\equiv -\frac{5}{3}\frac{t^L_{(0)}}{m^L}\left[m^A(1+\Delta^L)+\Delta^A+\mathcal{O}(\epsilon)\right]\, ,
\label{expansion}
\end{equation}
where we have defined $t^L_{(0)}$ to be the vev of $t^L$ in the factorised case. For future reference, we also define the leading-order vev of $t^A_{(0)}$ as well as the next-to-leading-order corrected vevs $t^L_{(1)}$ and $t^A_{(1)}$. We thus have
\begin{equation}
\begin{aligned}
&t^L_{(0)}\equiv\sqrt{\frac{-9Qm^L}{5M}}\ ,\qquad
&&t^A_{(0)}\equiv -\frac{5m^A}{3m^L}t^L_{(0)}\ ,\\
&t^L_{(1)}\equiv t^L_{(0)}\left(1+\Delta^L\right)\, ,\qquad &&t^A_{(1)}\equiv t^A_{(0)}-\frac{5}{3}\frac{t^L_{(0)}}{m^L}\left(\Delta^A+m^A\Delta^L\right)\, ,
\end{aligned}
\end{equation}
where ${\cal Q}$ scales like in the factorised case: ${\cal Q}\sim n^{2r}$. For the expansion \eqref{expansion} to make sense, we need the corrections to behave like $\Delta^L=\mathcal{O}(\epsilon)\sim n^{s-r}$ and $\Delta^A=\mathcal{O}(1)\sim\text{const.}$ Inserting the expansion inside the equations \eqref{saxfib} we find
\begin{equation}
\Delta^L=-\frac{\eta_{AB}c^Am^Bm^L}{2M}\, ,\qquad \Delta^A=\frac{4}{5}c^Am^L\, ,
\end{equation}
which indeed scale consistently.

This perturbative analysis thus suggests a mechanism to generate an infinite family of scale-separated vacua with scalings \eqref{newsc} and \eqref{newsc2} asymptotically at large $n$, thanks to the choice
\begin{equation}
m^A\sim n^{r-s}\, ,\quad m^L\sim \text{const.}\, ,\quad {\cal Q}\sim n^{2r}\, ,
\end{equation}
which this time only involves flux quanta allowed to scale, exactly as in the factorised case. Appendix \ref{ap:numerical} performs a numerical analysis of the 4d vacua equations for two specific elliptically fibered models, and confirms the existence of the infinite family of vacua as well as the robustness of the analytical results derived in this section.


\section{Stability}
\label{s:stability}

To study the perturbative stability of the vacua described in this paper, we follow the same approach as in \cite{Marchesano:2019hfb,Marchesano:2020uqz}. That is, we compute the Hessian matrix from the scalar F-term potential \eqref{eq:potentialgeom} in our branches of solutions \eqref{branches} under consideration. From there, to obtain the canonically normalised masses, we make use of the decomposition of the Kähler metric into a primitive and non-primitive part \cite{Marchesano:2019hfb,Marchesano:2020uqz}.

We will first display the generic expression of the Hessian, valid for any SU(3)-structure compactification within the branches of vacua \eqref{Ansatz1} and \eqref{branches}. With this general expression it is not possible to extract the eigenvalues in full generality. This task is however doable for the twisted factorised geometries $X_6 = (T^2 \tilde{\times} X_4)/\Gamma$ with $X_4 = T^2 \times T^2$ or $K3$ considered in sect.~\ref{sec:factorized}. In both cases the masses are the same and do not induce any perturbative instability. For the case of more general elliptic fibrations, we expect to recover a similar mass spectrum at large scaling parameter $n$ where the perturbative expansion described in sect.~\ref{pert} converges towards the factorised setup. We will refer to numerical results given in Appendix~\ref{ap:numerical} to confirm this expectation and conclude that our new infinite family of scale-separated vacua is perturbatively stable. We will eventually discuss aspects of stability at the non-perturbative level in sect.~\ref{nonpert}.

\subsection{Generic Hessian matrix}

The strategy to express the Hessian mirrors the one implemented in \cite[Appendix C]{Marchesano:2020uqz}. We thus proceed with the following steps:
\begin{itemize}
    \item From the generic scalar potential \eqref{eq:potentialgeom}  translated to Einstein frame and expressed in four-dimensional Planck units, one can compute double derivatives with respect to all the fields and this defines the Hessian matrix. 
    \item One then particularises these expressions to the branches of solutions under consideration by implementing the Ansatz \eqref{proprhog} and the definition of the branches \eqref{branches}. It is already useful at this point to decompose the Kähler metric into its primitive and non-primitive parts, both for the Kähler and complex-structure sectors. We have
    \begin{equation}
    \begin{aligned}
    &g_{ab}^{\rm NP}=\frac{1}{12}\partial_aK\partial_bK\, , &&g_{ab}^{\rm P}=\frac{1}{6}\partial_aK\partial_bK-\frac{3}{2}\frac{\cK_{ab}}{\cK}\,  &&g^{ab}_{\rm NP}=\frac{4}{3}t^at^b\, , &&g^{ab}_{\rm P}=-\frac{2}{3}\left(\cK\cK^{ab}-t^at^b\right)\, ,\\
    &c_{\mu\nu}^{NP}=\frac{1}{16}\frac{G_\mu G_\nu}{G^2}\, , &&c_{\mu\nu}^{\rm P}=\frac{3}{16}\frac{G_\mu G_\nu}{G^2}-\frac{1}{4}\frac{G_{\mu\nu}}{G}\, ,&& c^{\mu\nu}_{\rm NP}=u^\mu u^\nu\, , &&c^{\mu\nu}_{\rm P}=\frac{1}{3}u^\mu u^\nu-4GG^{\mu\nu}\, ,
    \end{aligned}
    \end{equation}
   where the subscripts and superscripts P and NP stand for the primitive and non-primitive parts of the metrics and their inverses, $G\equiv e^{-K_Q}$, $G_\mu\equiv\partial_{u^\mu} G$ and $G_{\mu\nu}\equiv\partial_{u^\mu} \partial_{u^\nu} G$.

    At this point, one can realise that the matrix is block diagonal with the saxions decoupled from the axions.
    \item For the eigenvalues of the matrix to be mapped to the canonically normalised masses, the Hessian should be expressed in an orthonormal basis which splits into vectors along the (one dimensional) non-primitive subspaces of the Kähler metrics and the ($h^{1,1}_-$- and $h^{2,1}$-dimensional) primitive subspaces \cite{Marchesano:2019hfb,Marchesano:2020uqz}. The unit-norm vectors along the non-primitive spaces are denoted $\hat\xi$, $\hat b$, $\hat u$ and $\hat t$ for the axions and saxions respectively. On the other hand, the unit-norm vectors spanning the primitive spaces are written $\hat\xi_{\hat\mu}$, $\hat b_{\hat\alpha}$, $\hat u_{\hat\mu}$ and $\hat t_{\hat\alpha}$ with $\hat\mu\in\{1,\dots,h^{2,1}-1\}$ and $\hat\alpha\in\{1,\dots,h^{1,1}_--1\}$. The change of basis thus reads
    \begin{equation}
    \label{basis}
    (\xi^\mu,b^a)\longrightarrow(\hat\xi,\hat b,\hat\xi_{\hat\mu},\hat b_{\hat\alpha})\quad\text{ and }\quad (u^\mu,t^a)\longrightarrow (\hat u,\hat t,\hat u_{\hat\mu},\hat t_{\hat\alpha})\, .
    \end{equation}
\end{itemize}

The implementation of all these steps is a rather involved process and we will simply display the result. We denote $A$ the axionic submatrix and $S$ the saxionic one. We find

\bes
\label{H}
\begin{align}
\label{eq:Haxions}
&A=\begin{pmatrix}
16Q^2\left(4-\frac{\Sigma}{\sigma^2}\right) & -\frac{32Q^2}{\sqrt{3}}\left(6\delta-1-\frac{\Sigma}{\sigma^2}\right) & 0 & -24\delta QJ_{\hat\alpha}+\mathcal{J}_{\hat\alpha}\\
-\frac{32Q^2}{\sqrt{3}}\left(6\delta-1-\frac{\Sigma}{\sigma^2}\right) & \frac{16Q^2}{3}\left(11+6\delta-\frac{4\Sigma}{\sigma^2}\right) & 0 & \frac{2}{\sqrt{3}}\left(24\delta QJ_{\hat\alpha}-\mathcal{J}_{\hat\alpha}\right)\\
0 & 0 & 0 & 0\\
-24\delta QJ_{\hat\alpha}+\mathcal{J}_{\hat\alpha} &  \frac{2}{\sqrt{3}}\left(24\delta QJ_{\hat\alpha}-\mathcal{J}_{\hat\alpha}\right) & 0 & 12(-1)^{\delta+1} QJ_{\hat\alpha\hat\beta}-\mathcal{J}_{\hat\alpha\hat\beta}+6J_{\hat\alpha}J_{\hat\beta}
\end{pmatrix}e^K,\\\label{eq:Hsaxions}
&S=\begin{pmatrix}
16Q^2\left(7-\frac{\Sigma}{\sigma^2}\right) & \frac{32Q^2}{\sqrt{3}}\left(2+\frac{\Sigma}{\sigma^2}\right) & 0 & -8QJ_{\hat\alpha}+\mathcal{J}_{\hat\alpha}\\
\frac{32Q^2}{\sqrt{3}}\left(2+\frac{\Sigma}{\sigma^2}\right) & \frac{16Q^2}{3}\left(19-4\frac{\Sigma}{\sigma^2}\right) & 0 & \frac{2}{\sqrt{3}}\left(8QJ_{\hat\alpha}-\mathcal{J}_{\hat\alpha}\right)\\
0 & 0 & -16Q^2 & 0\\
-8QJ_{\hat\alpha}+\mathcal{J}_{\hat\alpha} &  \frac{2}{\sqrt{3}}\left(8QJ_{\hat\alpha}-\mathcal{J}_{\hat\alpha}\right) & 0 & -16Q^2\delta_{\hat\alpha\hat\beta}-4QJ_{\hat\alpha\hat\beta}-\mathcal{J}_{\hat\alpha\hat\beta}+6J_{\hat\alpha}J_{\hat\beta}
\end{pmatrix}e^K,
\end{align}
\ees
where $\Sigma\equiv e^{-K_K}\sigma_a\cK^{ab}\sigma_b$ and similarly $\mathcal{J}_{ab}\equiv e^{-K_K}J_{ac}\cK^{cd}J_{db}$. The quantities with hatted indices are obtained by contractions with $\hat t^a_{\hat\alpha}$. For example, $J_{\hat\alpha b}\equiv J_{ab} \hat t^a_{\hat\alpha}$ and the single-hatted-index matrices $J_{\hat\alpha}$ and $\mathcal{J}_{\hat\alpha}$ are obtained by further contraction with $t^b$. The quantity $\delta$ is $0$ or $1$ respectively at vacua that are supersymmetric or not. Notice that the saxionic mass matrix does not depend on $\delta$ and it is thus identical in the SUSY and non-SUSY branches. The third row/column associated to $\hat\xi_{\hat\mu}$ and $\hat u_{\hat\alpha}$ is understood to be $h^{2,1}$-dimensional and diagonal.

Note that these expressions for the Hessian are only implicit in the sense that it is not possible to have knowledge about the canonically normalised basis in full generality. We can however uncover a universal spectrum for simple geometries as we will see now.

\subsection{Twisted factorised geometries}
\label{ss: stability in factorised cases}

Let us investigate the Hessian for the twisted factorised geometries $X_6 = (T^2 \tilde{\times} X_4)/\Gamma$ considered in sect.~\ref{sec:factorized} with $X_4 = T^2 \times T^2$ or K3. We first focus on the orbifold untwisted sector for each choice of  compactification, and then comment on the orbifold twisted sector. As we argue, we do not expect perturbative instabilities to arise in such a twisted sector, although this will in general depend on the specific choice of $\Gamma$. 

\paragraph{\bm $T^6$ with metric fluxes\\}

Let us consider the case $X_4 = T^2 \times T^2$ that corresponds to a factorised $T^6$ with metric fluxes, and an abelian orbifold $\Gamma$ such that $h^{1,1}_{\rm untw}(T^6/\Gamma) = 3$. Then one can explicitly express the three basis vectors $(\hat t,\hat t_{\hat\alpha})$ (the vectors $(\hat b,\hat b_{\hat\alpha})$ are then the same, acting on the proper axionic subspace). The only non-vanishing intersection number is $\cK_{123}$ such that the Kähler sector metric reads
\begin{equation}
g_{ab}^{\rm P}=\frac{1}{12}\begin{pmatrix}
\frac{2}{(t^1)^2} & -\frac{1}{t^1t^2} & -\frac{1}{t^1t^3}\\
-\frac{1}{t^1t^2} & \frac{2}{(t^2)^2} & -\frac{1}{t^2t^3}\\
-\frac{1}{t^1t^3} & -\frac{1}{t^2t^3} & \frac{2}{(t^3)^2}
\end{pmatrix},\qquad
g_{ab}^{\rm NP}=\frac{1}{12}\begin{pmatrix}
\frac{1}{(t^1)^2} & \frac{1}{t^1t^2} & \frac{1}{t^1t^3}\\
\frac{1}{t^1t^2} & \frac{1}{(t^2)^2} & \frac{1}{t^2t^3}\\
\frac{1}{t^1t^3} & \frac{1}{t^2t^3} & \frac{1}{(t^3)^2}
\end{pmatrix},
\end{equation}
from which we deduce the following orthonormal basis:
\begin{equation}
\hat t=\frac{2}{\sqrt{3}}(t^1,t^2,t^3)\, ,\quad \hat t_{\hat 1}=\sqrt{2}(-t^1,0,t^3)\, \quad t_{\hat 2}=\sqrt{\frac{2}{3}}(-t^1,2t^2,-t^3)\, .
\end{equation}
With this explicit basis, with $\sigma_a=(1,0,0)$ to match the structure \eqref{Kfact} and with the saxionic vevs given by \eqref{vevsfact}, we find the following universal behaviour for the Hessian:

\begin{equation}
A=8\left(\begin{array}{ccccc}
  16 & -4\sqrt{3}(1+2\delta) & 0 & -12\sqrt{2}(1-\delta) & -4\sqrt{6}(1-\delta)\\
-4\sqrt{3}(1+2\delta) & 2(9+2\delta) & 0 & 8\sqrt{6}(1-\delta) & 8\sqrt{2}(1-\delta)\\
0 & 0 & 0 & 0 & 0\\
-12\sqrt{2}(1-\delta) & 8\sqrt{6}(1-\delta) & 0 & 49+9(-1)^\delta & 8\sqrt{3}(1+(-1)^\delta)\\
-4\sqrt{6}(1-\delta) & 8\sqrt{2}(1-\delta) & 0 & 8\sqrt{3}(1+(-1)^\delta) & 33-7(-1)^\delta \end{array}\right)  e^K\mathcal{Q}^2\, ,
\end{equation}

\begin{equation}
S=\left(
\begin{array}{ccccc}
 176 & -\frac{64}{\sqrt{3}} & 0 & -64 \sqrt{2} & -64 \sqrt{\frac{2}{3}} \\
 -\frac{64}{\sqrt{3}} & \frac{560}{3} & 0 & 128 \sqrt{\frac{2}{3}} & \frac{128
   \sqrt{2}}{3} \\
 0 & 0 & -16 & 0 & 0 \\
 -64 \sqrt{2} & 128 \sqrt{\frac{2}{3}} & 0 & 400 & \frac{256}{\sqrt{3}} \\
 -64 \sqrt{\frac{2}{3}} & \frac{128 \sqrt{2}}{3} & 0 & \frac{256}{\sqrt{3}} &
   \frac{688}{3} \\
\end{array}
\right)e^K\mathcal{Q}^2\, .
\end{equation}

To get the masses, we need to divide the eigenvalues by a factor two and it is also useful to normalise the spectrum with the Breitenlohner-Freedman bound $m_{\rm BF}$.  In Planck units, it is given by $m_{\rm BF}^2\equiv -3/4|V|$ which in our setup translates into $m_{\rm BF}^2=-9e^{K}{\cal Q}^2$. We obtain the following masses $m_{\rm s}^2$ for the saxions and $m_{\rm a}^2$ for the axions:
\begin{equation}
\label{universal_spectrum}
\frac{m_{\rm s}^2}{|m_{\rm BF}|^2}=\left(\frac{280}{9},8,8,8,-\frac{8}{9}\right)\, ,\quad\begin{cases}
\text{SUSY: }  &\frac{m_{\rm a}^2}{|m_{\rm BF}|^2}=(\frac{352}{9},\frac{40}{9},\frac{40}{9},\frac{40}{9},0)\\
\text{non-SUSY: }  &\frac{m_{\rm a}^2}{|m_{\rm BF}|^2}=(\frac{160}{9},\frac{160}{9},\frac{160}{9},-\frac{8}{9},0)\
\end{cases}
\end{equation}
The saxionic tachyon has multiplicity $h^{2,1}_{\rm untw}(T^6/\Gamma) \leq 3$ and is just above the BF bound. On the other hand, the axionic massless mode has the same multiplicity and the tachyon in the non-SUSY branch is also above the BF bound. There is thus no instability induced at the perturbative level. Note that this spectrum matches the universal one found in the T-dual DGKT-CFI setup \cite{Marchesano:2019hfb}. We interpret this exact matching as a consequence of the factorised structure for the triple intersection numbers, which allows us to implement T-duality on a K\"ahler modulus that does not mix kinematically with the rest of the (untwisted) K\"ahler sector. 

\paragraph{\bm $T^2 \times$ K3 with metric fluxes\\}

Now we consider the slightly more involved example that arises when we take $X_6 = (T^2 \tilde{\times} X_4)/\Gamma$  with $X_4$ a K3 manifold. One may attempt to work directly with \eqref{eq:Haxions} and \eqref{eq:Hsaxions}, however, there is a simpler way to build the canonically normalised masses and recover a universal behaviour of the Hessian by taking advantage of the factorised structure of  metric and the fact that the metric fluxes run along the fibre. We thus treat separately the direction of the fibre and only split the base sector of the fibration in its primitive and non primitive components. The metric in this case is of the form
\begin{equation}
    g_{ab}=\left(\begin{array}{cc}
        \frac{1}{4(t^L)^2} &0  \\
        0 & \frac{\eta_A\eta_B}{\eta^2}-\frac{1}{2}\frac{\eta_{AB}}{\eta}
    \end{array}\right)\,,\qquad 
    g^{ab}=\left(\begin{array}{cc}
        4(t^L)^2 &0  \\
        0 & 4t^At^B-2\eta^{AB}\eta
    \end{array}\right)\,,
\end{equation}
where $A,B$ go from $1$ to $h^{1,1}_-(X_4)$ and $\eta_{AB}$ are the intersection numbers of the $X_4$ base. We can split the base part of the metric in its primitive and non-primitive sectors, obtaining
\begin{align}
    g_{AB}^{\rm NP}=\frac{1}{2}\frac{\eta_A\eta_B}{\eta^2}\,,\qquad  & g_{AB}^{\rm P}=\frac{1}{2}\left(\frac{\eta_A\eta_ B}{\eta^2}-\frac{\eta_{AB}}{\eta}\right)\, ,\\
    g^{AB}_{\rm NP}=2t^At^B\,,\qquad & g^{AB}_{\rm P}=2t^At^B-2\eta^{AB}\eta\, .
\end{align}

We introduce a convenient basis $(\hat{t}^L,\hat{t}, \hat{t}_{\hat{\alpha}})$, with $\hat{t}^L$ defined along the fibre, $\hat{t}$ in the non-primitive sector of the base and $\hat{t}_{\hat{\alpha}}$ in the primitive sector of the base. Since we want all of them to be orthonormal we choose 
\begin{equation}
    \hat{t}^L= 2(t^L,0,\dots)\,,\qquad \hat{t}=\sqrt{2}(0,t^1,\dots)\, ,
\end{equation}
and leave the conditions for the primitive sector of the base implicit. In the basis ($\hat{u},\hat{u}_{\hat{\mu}},\hat{t}^L,\hat{t}, \hat{t}_{\hat{\alpha}})$  the Hessian becomes
\begin{equation}
    A=16\left(\begin{array}{ccccc}
        8\mathcal{Q}^2 & 0 & 6(-1+2\delta)  & -6\sqrt{2} & 0 \\
        0 & 0 & 0 & 0 & 0\\
        6(-1+2\delta) & 0 & 34 & (-12+9\delta)\sqrt{2} & 0\\
        -6\sqrt{2} & 0 & (-12+9\delta)\sqrt{2} & (29-15\delta) &0 \\
        0 &0 &0 &0 & 5+15\delta
    \end{array}\right)\mathcal{Q}^2 e^K\, ,
\end{equation}

\begin{equation}
   S=16\left(\begin{array}{ccccc}
       11 & 0 & 4 & -4\sqrt{2} & 0 \\
        0 & -1 & 0 & 0 & 0\\
        4 & 0 & 17 & -8\sqrt{2} & 0\\
        -4\sqrt{2}& 0 & -8\sqrt{2} & 25 &0 \\
        0 &0 &0 &0 & 9
    \end{array}\right)\mathcal{Q}^2 e^K\, ,
\end{equation}
where the entries associated to the second component of the basis have multiplicity $h^{2,1}_{\rm untw}(T^2 \times K3/\Gamma)$ and the ones associated to the fifth component have multiplicity $h_-^{1,1}(X_4)-1$. In particular the entry $(5,5)$ corresponds to a diagonal matrix with shape $(h_-^{1,1}(X_4)-1)\times (h_-^{1,1}(X_4)-1)$.

To get the masses, we divide the eigenvalues by a factor two and normalise the spectrum with the Breitenlohner-Freedman bound $m_{\rm BF}^2=-9e^{K}{\cal Q}^2$ as in the previous example. We obtain the following masses $m_{\rm s}^2$ for the saxions and $m_{\rm a}^2$ for the axions:
\begin{equation}
\label{universal_spectrum K3xT2}
\frac{m_{\rm s}^2}{|m_{\rm BF}|^2}=\left(\frac{280}{9},8,8,8,-\frac{8}{9}\right)\, ,\quad\begin{cases}
\text{SUSY: }  &\frac{m_{\rm a}^2}{|m_{\rm BF}|^2}=(\frac{352}{9},\frac{40}{9},\frac{40}{9},\frac{40}{9},0)\\
\text{non-SUSY: }  &\frac{m_{\rm a}^2}{|m_{\rm BF}|^2}=(\frac{160}{9},\frac{160}{9},\frac{160}{9},-\frac{8}{9},0)\
\end{cases}
\end{equation}
where the multiplicities of the masses are $(1,h_-^{1,1}(X_4)-1,1,1,h^{2,1}_{\rm untw}(T^2 \times K3/\Gamma))$. As we can see, the mass spectrum is the same as in \eqref{universal_spectrum}, up to multiplicity. Again, this is to be expected from the relatively simple setting, that displays a factorisation of the triple intersection numbers. These allow us to link our setup to DGKT-like vacua via T-duality on a K\"ahler modulus kinematically decoupled from the rest of the untwisted K\"ahler sector, which presumably explains the exact matching with the universal spectrum found in \cite{Marchesano:2019hfb}.

\paragraph{\bm Orbifold twisted sector\\}

Let us now turn to potential instabilities in the closed-string twisted sector localised at fixed loci of the orbifold group $\Gamma$. By assumption, such fields have vanishing vev and do not enter the flux superpotential, so we can describe them as a perturbation of the untwisted K\"ahler potential, as usually done for the chiral sector of a compactification \cite{Brignole:1997wnc}:
\be
K =  K_K^{\rm untw} + K_Q^{\rm untw} + \tilde{K}_{\a\bar{\b}} (H_m, \bar{H}_m) C^\a \bar{C}^{\bar{\b}} + \cdots 
\label{Ktwist}
\ee
Here $H_m = \{ T^a, U^\mu\}$ represent the untwisted sector, and $C^\a$ the closed string fields of the twisted sector. For many orbifolds quotients the K\"ahler metrics of the twisted sector are diagonal
\be
\tilde{K}_{\a\bar{\b}} (H_m, \bar{H}_m)= \delta_{\a\bar{\b}} \tilde{K}_\a (H_m, \bar{H}_m)\, ,
\ee
implying that we can use the following formula for the flux-induced mass on each twisted field \cite{Brignole:1997wnc}, expressed in Planck units:
\be
m_\a^2 =  V + e^K|W|^2 - e^K F^n \bar{F}^{\bar{m}} \p_{H^n}     \p_{\bar{H}^{\bar{m}}} \log K_\a\, ,
\label{softdiag}
\ee
with the rhs evaluated at the vacuum. For the supersymmetric branch of vacua, where F-terms vanish, we find
\be
m_\a^2 = - \frac{2}{3} |V|_{\rm vac} = -\frac{8}{9} |m_{\rm BF}|^2\, ,
\ee 
matching the mass spectrum of the untwisted complex structure fields that do not enter the superpotential. For the non-supersymmetric branch \eqref{nosusybranch} we need to make use of the F-term structure \eqref{solsfmax}, \eqref{Ftermnosusy}. From here we obtain that the universal contribution to \eqref{softdiag} reads
\be
V|_{\rm vac} + e^K|W|^2_{\rm vac} =  13e^K \mathcal{Q}^2 \, ,
\ee
which is positive. To generate a perturbative instability in the twisted sector, the non-universal contribution to \eqref{softdiag} should therefore be negative and large enough to violate the BF bound. To evaluate this second contribution one needs to specify the choice of $\Gamma$. To get an estimate, let us consider the K\"ahler metrics that arise in toroidal orbifolds of the form $T^6/\Gamma$, with $\Gamma$ an Abelian orbifold:
\be
\tilde{K}_{\a\bar{\b}} \propto \delta_{\a\bar{\b}} \prod_{a=1}^{h^{1,1}_{\rm untw}} \left(T^a-\bar{T}^{\bar{a}}\right)^{n_\a^a} \prod_{\mu=1}^{h^{2,1}_{\rm untw}} \left(U^\mu-\bar{U}^{\bar{\mu}}\right)^{l_\a^\mu} \, ,
\ee
where the rational numbers $n_\a^a$, $l_\a^\mu$ are the modular weights of the orbifold singularity \cite{Ibanez:1992hc,Bailin:1992he}. From here one can deduce that the non-universal contribution to the twisted masses is
\be
- e^K F^a \bar{F}^{\bar{a}} \p_a \p_{\bar{a}} \log K_\a|_{\rm vac} =  9 L_\a e^K  \mathcal{Q}^2\, , \qquad L_\a = \sum_{a=1}^{h^{1,1}_{\rm untw}} n_\a^a + \sum_{\mu=1}^{h^{2,1}_{\rm untw}} l_\a^\mu\, .
\ee
The number $L_\a$ is typically negative, lowering the mass for the twisted field, but as long as it takes values above $-22/9$, as is often the case, no instability will be induced. Finally, notice that this result should be reconsidered if the twisted fields enter the flux superpotential, as further contributions to the mass terms would then be induced by a Giudice-Masiero mechanism.

\subsection{Elliptic fibrations}

For non-trivial elliptic fibrations, one can neither explicitly express the primitive/non-primitive basis as in the factorised torus case nor exploit the block diagonal aspect of the metric between the fibre and the base components as in the $T^2\tilde\times X_4$ setup. Moreover, unlike in our previous analysis, now we observe deviations from the universal DGKT-like mass spectrum \eqref{universal_spectrum}. This could be interpreted from the fact that, in the T-dual DGKT-like picture, there is a K\"ahler modulus that is small and that mixes kinematically with all the remaining K\"ahler moduli. As such, to fully account for T-duality at the level of the 4d EFT one should incorporate to the K\"ahler potential \eqref{KK} the associated leading curvature corrections, which were neglected in the large-volume setup of \cite{Marchesano:2019hfb}. However, we still expect from the perturbative analysis developed in sect.~\ref{pert} to recover the universal mass spectrum asymptotically at large $n$, since in this limit the said K\"ahler modulus decouples kinematically from the rest. As such, our infinite family of scale separated vacua is safe from perturbative instabilities. 

To check these expectations we explored numerically two specific examples of non-trivial elliptic fibrations. We generated vacua along our infinite family with increasing $n$ and computed numerically, for each one of them, the Hessian matrix from the scalar potential \eqref{eq:potentialgeom}. At a given vacuum with completely determined numerical vevs, it is then easy to perform the change of basis in order to get the canonically normalised masses. Alternatively, instead of starting from the potential, we could have evaluated directly the expressions \eqref{eq:Haxions} and \eqref{eq:Hsaxions}. However, the former strategy does not rely on the heavy computations required to obtain the analytical Hessian and is thus safer in this sense. Moreover, this way we could check that the analytical expressions match the Hessian obtained from the potential and we indeed observed a perfect agreement. These results are presented in more details in Appendix~\ref{ap:numerical} and the bottom line of the analysis is pictured in figs.~\ref{fig:masses1} and \ref{fig:masses2}. From these graphics, we see that the universal mass spectrum \eqref{universal_spectrum} is indeed recovered asymptotically at large $n$ (small $\epsilon$) in the two specific compactifications under study, which strongly supports the perturbative stability of this new infinite family of scale separated vacua.

\subsection{Non-perturbative stability}
\label{nonpert}

At the non-perturbative level, one should consider the stability of these vacua against membrane nucleation. In many aspects, the discussion parallels the one already performed in the literature for DGKT-like vacua, see \cite{Aharony:2008wz,Narayan:2010em,Marchesano:2021ycx,Casas:2022mnz,Marchesano:2022rpr}, in particular for those vacua related by T-duality. 

For the supersymmetric branch of vacua, the main question is whether our 4d solution based on a smeared sources uplifts to a fully-fledged type IIA string compactification with four preserved supercharges. If it does, then non-perturbative stability follows from the general analysis of \cite{Giri:2021eob}. As already mentioned, at the level of 10d supergravity backgrounds one expects source-localisation to be achieved as soon as one considers deformations to SU(3)$\times$SU(3) metric background on $X_6$, as it happens for DGKT-like vacua in \cite{Junghans:2020acz,Marchesano:2020qvg}. In particular, since such a SU(3)$\times$SU(3) deformation can be understood as a first-order correction in a $g_s$ perturbative expansion, one expects that T-duals of DGKT-like vacua enjoy the same property. Indeed, the same strategy has been applied in \cite{Cribiori:2021djm} to the factorised $T^6$ with metric fluxes, obtaining similar results. Computing further corrections in this perturbative expansion would also require to take $\alpha'$ corrections into account, presumably taking us away from the  10d supergravity description. While this remains an open problem,  one would expect that if the perturbative expansion in $g_s$ and $\a'$ can be completed for DGKT-like vacua, the same can be done for those vacua related by T-duality, and the other way around. The same cannot be said for possible supersymmetry breaking effects occurring at the non-perturbative level in $g_s$ or $\a'$, as recently proposed in \cite{talkmiguel}, because T-duality is not guaranteed to hold at this level \cite{Aspinwall:1999ii}. Therefore, if a supersymmetry-breaking effect generating an instability for perturbatively supersymmetric DGKT vacua was found, the analysis should be repeated in the present context.  

For the non-supersymmetric branch of vacua, a guiding principle to search for non-perturbative instabilities is the sharpened Weak Gravity Conjecture applied to 4d membranes \cite{Ooguri:2016pdq}. More precisely, one may apply the analysis of \cite{Marchesano:2021ycx} to test the stability of 4d membranes in a Poincar\'e patch of AdS$_4$ parallel to its boundary, in the present context. Using the results of section~\ref{s:general} we may do so at the level of smeared-source backgrounds, focusing for concreteness on the sub-branch \eqref{Ansatz1}. From \eqref{eq: scalar potential vacuum new ansatz} one first deduces the following relation for  the AdS$_4$ radius $R$ measured in the 10d string frame
\begin{equation}
    \frac{\ell_s}{R} = \frac{g_s}{2{\rm Vol}_{X_6}} |{\cal Q}| \, .
\end{equation}
Then, from the smeared internal fluxes threading $X_6$ one deduces the following fluxes that translate into 4d four-form fluxes upon dimensional reduction: 
\begin{equation}
   G_4 = \frac{3\iota\iota_{\cal Q}}{Rg_s} {\rm vol}_4\, , \qquad G_8 =  \frac{3\iota_{\cal Q}}{Rg_s} {\rm vol}_4 \wedge \oh J^2 + g_s^{-1}  {\rm vol}_4 \wedge \dd\im \Omega \, ,
\end{equation}
where $\rho_0\equiv-\iota \frac{3}{2}\mathcal{Q}$ and $\iota_{\cal Q} \equiv {\rm sign }\, \mathcal{Q}$. Such four-form fluxes support 4d membranes that come from (anti-)D2-branes point-like in $X_6$ and from  (anti-)D6-branes wrapping calibrated four-cycles of $X_6$, respectively. Following \cite{Marchesano:2021ycx}, one can see that upon choosing the appropriate orientation one recovers 
\begin{equation}
    Q_{\rm D2/D6} = T_{\rm D2/D6}\, ,
\end{equation}
and these objects are marginally stable, both for the SUSY and non-SUSY branches in \eqref{branches}.  This is to be expected, as upon two T-dualities these objects are mapped to (anti-)D4-branes wrapping holomorphic two-cycles in DGKT-like vacua.\footnote{Bionic configurations of D8-branes in DGKT-like vacua \cite{Marchesano:2021ycx} are mapped by T-duality to D6-branes on four-chains of $X_6$ with a non-trivial boundary, on which space-time filling D6-branes wrapping trivial three-cycles end.} It would be interesting to check whether this equality is still maintained when including the first-order corrections to the 10d smeared background, as it happens for DGKT-like vacua.

\bigskip
\bigskip

\centerline{\bf  Acknowledgments}

\vspace*{.5cm}

We thank Alberto Castellano, Luca Melotti, Miguel Montero, Raffaelle Savelli, \'Angel Uranga, Irene Valenzuela and Max Wiesner for discussions.  This work is supported through the grants EUREXCEL$\_$03 funded by CSIC, and CEX2020-001007-S and PID2021-123017NB-I00, funded by MCIN/AEI/10.13039/501100011033 and by ERDF A way of making Europe. R. C. is supported through the JAE Intro grant JAEINT$\_21\_$02330. D. P. is supported through the grant FPU19/04298 funded by MCIN/AEI/10.1\-3039/501100011033 and by ESF Investing in your future, and would like to thank CERN-TH for hospitality during the development of this work.


\appendix

\section{Numerical analysis}
\label{ap:numerical}

In this appendix, we investigate numerically two specific elliptically fibered geometries. The examples that we consider are (the T-duals of) the Calabi-Yau 3-folds obtained by resolving the singularities of a degree $18$ hypersuface of $\mathbb{P}^4_{(1,1,1,6,9)}$  and a degree $24$ hypersurface of $\mathbb{P}^5_{(1,1,2,8,12)}$, which have two and three Kähler parameters respectively. The purpose of these numerical explorations is fourfold:
\begin{itemize}
\item It allows us to confirm, at the 4d level, the existence of the infinite family of scale-separated vacua for elliptically fibered geometries, which is described in the paper only through a perturbative expansion in section \ref{pert}.
\item It is also a way to check that the analytics of the perturbative approximation are consistent.
\item The perturbative stability of the vacua can only be assessed analytically for the twisted factorised torus or the trivial fibration $K3\times T^2$  with metric fluxes, following the computations developed in section \ref{ss: stability in factorised cases}. Numerics are a way to address the more general elliptically fibered case in specific examples to see how perturbative stability is affected.
\item A numerical computation of the Hessian starting directly from the scalar potential \eqref{eq:potentialgeom} allows us to crosscheck that the fully generic analytical expressions given by eqs.~\eqref{eq:Haxions} and \eqref{eq:Hsaxions} are indeed correct.
\end{itemize}

To generate numerical vacua, we choose the flux quanta to be given by their scalings \eqref{scalingsfib} (and \eqref{scalings_Q} to correctly scale ${\cal Q}$) with unit coefficients and appropriate signs for solutions to exist depending on the branch under consideration (supersymmetric or not). We  then vary $n$ to scan a wide range for the expansion parameter $\epsilon\sim n^{s-r}$. For the sake of clarity in the graphics, we choose the coefficients $r$ and $s$ to be $r=2$ and $s=1$ such that the solutions do not converge too quickly. Note that these values are both in agreement with flux quantification and allow for weak coupling as discussed below \eqref{gs}. A bigger difference $r-s$ would only make things converge faster. The primitive/non-primitive decomposition in the complex-structure sector is assumed to hold such that we only implement two complex-structure directions regardless of the actual Hodge number $h^{2,1}$. This is enough to evaluate the two masses (axion and saxion) corresponding to the non-primitive subspace and the remaining ones, along the primitive subspace, that are degenerate. 

\subsection[The $\mathbb{P}^4_{(1,1,1,6,9)}$ model]{\bm The $\mathbb{P}^4_{(1,1,1,6,9)}$ model}

This two-parameter elliptically fibered geometry over a base $\mathbb{P}^2$ has the following triple intersection numbers, where $L$ denotes the direction along the fibre and $i=1$ the other direction:
\begin{equation}
\cK_{LLL}=9\ ,\quad \cK_{LL1}=3\ ,\quad \cK_{L11}=1\ .
\end{equation}

\begin{figure}[ht!]
\begin{center}
\includegraphics[scale=0.27]{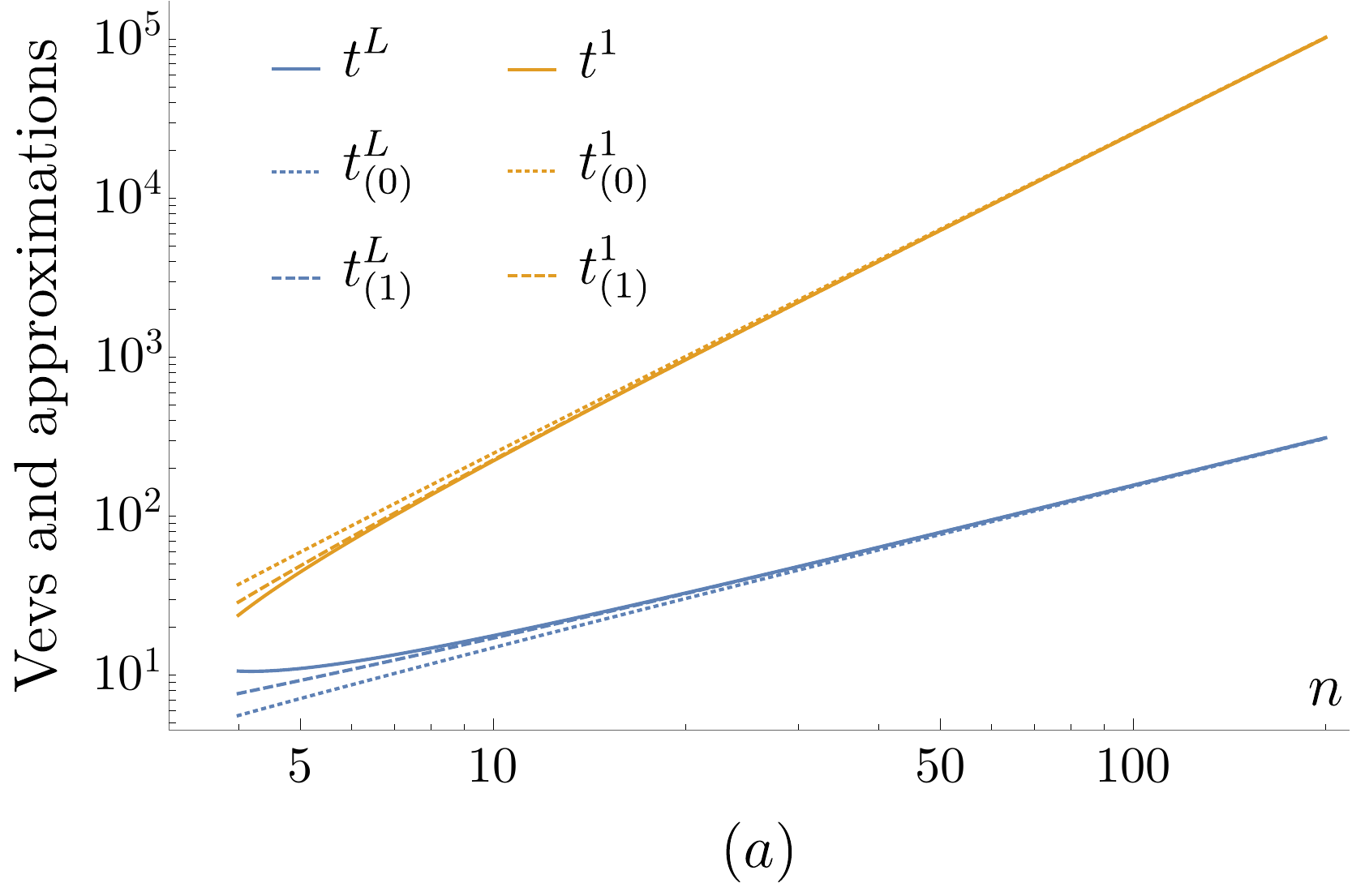}
\quad
\includegraphics[scale=0.27]{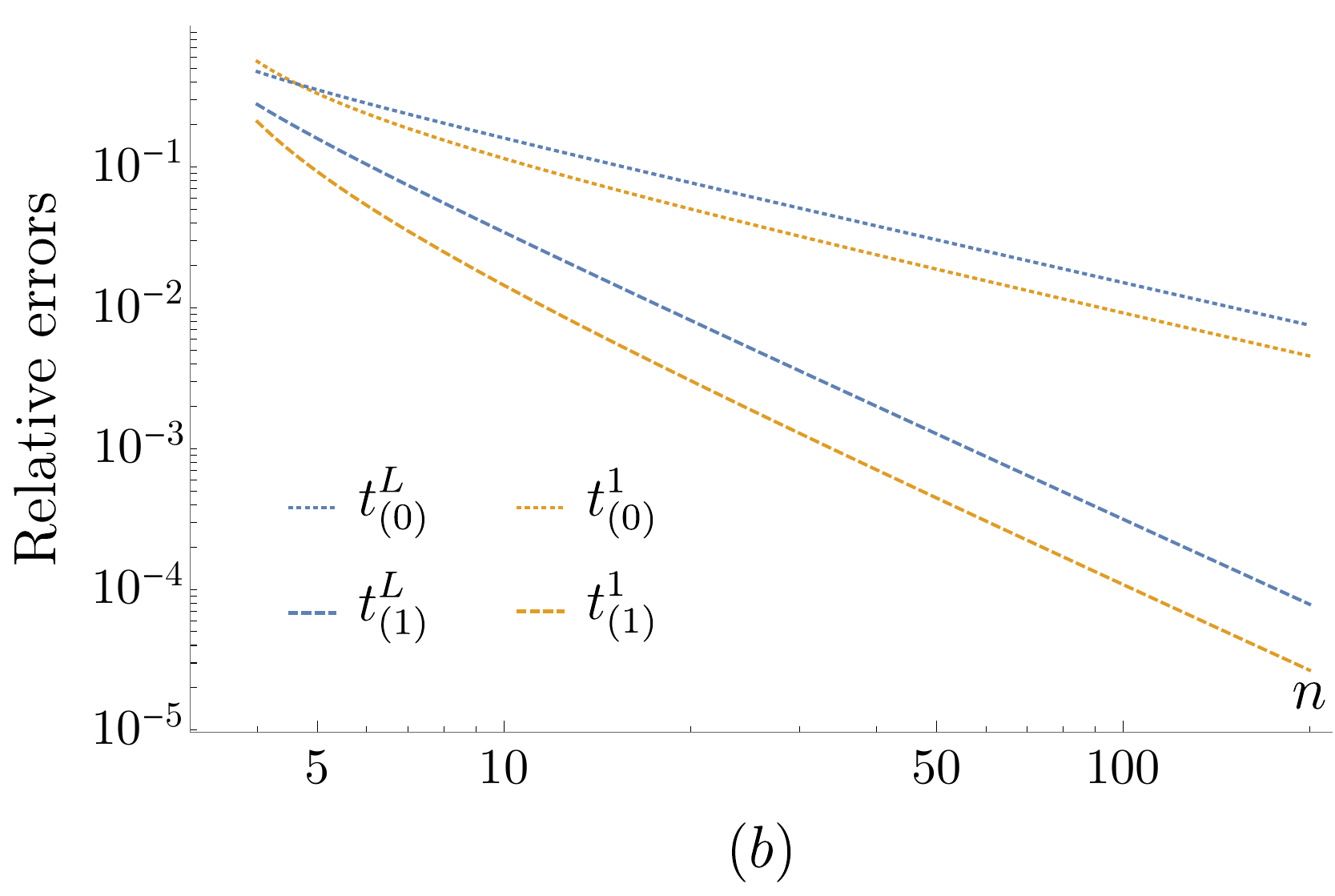}
\caption{(a) Vacuum expectation values of the saxions $t^L,t^1$ in the two-parameter model $\mathbb{P}^4_{(1,1,1,6,9)}$ and their leading-order and next-to-leading order approximations $t^L_{(0)}$, $t^1_{(0)}$, $t^L_{(1)}$ and $t^1_{(1)}$. (b) The relative error of these approximations with respect to the actual solution. The plots have been generated in the supersymmetric branch but similar behaviours are observed for the non-supersymmetric vacua.}
\label{fig:vevs1}
\end{center}
\end{figure}

Fig.~\ref{fig:vevs1}(a) shows the exact solutions $t^L$ and $t^1$ of the vacuum equations \eqref{saxfib} as well as the leading order and next-to-leading order approximations $t^L_{(0)}$, $t^1_{(0)}$, $t^L_{(1)}$ and $t^1_{(1)}$ respectively, with respect to $n$ in the range $[4,200]$. We observe that the leading order expressions are indeed good approximations of the actual solution and that the next-to-leading order one is even better, as expected. To make these statements more precise, fig.~\ref{fig:vevs1}(b) displays the relative error of the approximations with respect to the true solution. These errors scale exactly as they should in our setup: $n^{s-r}=n^{-1}$ for the leading approximation and $n^{2(s-r)}=n^{-2}$ for the next-to-leading one, which provides a strong check of the validity of our analytical expansion. This also confirms the existence of the infinite family of scale-separated vacua characterised in this paper.

For each vacua, we evaluate the Hessian directly from the scalar potential given by \eqref{eq:potentialgeom} and we express it in the primitive/non-primitive basis defined by \eqref{basis}. Note that going to this basis is hard in full generality but once we sit at a vacuum, the basis can easily be found numerically. A comparison of what is obtained with a direct implementation of the analytical expressions \eqref{eq:Haxions} and \eqref{eq:Hsaxions} shows a perfect agreement.

\begin{figure}[ht!]
\begin{center}
\includegraphics[scale=0.27]{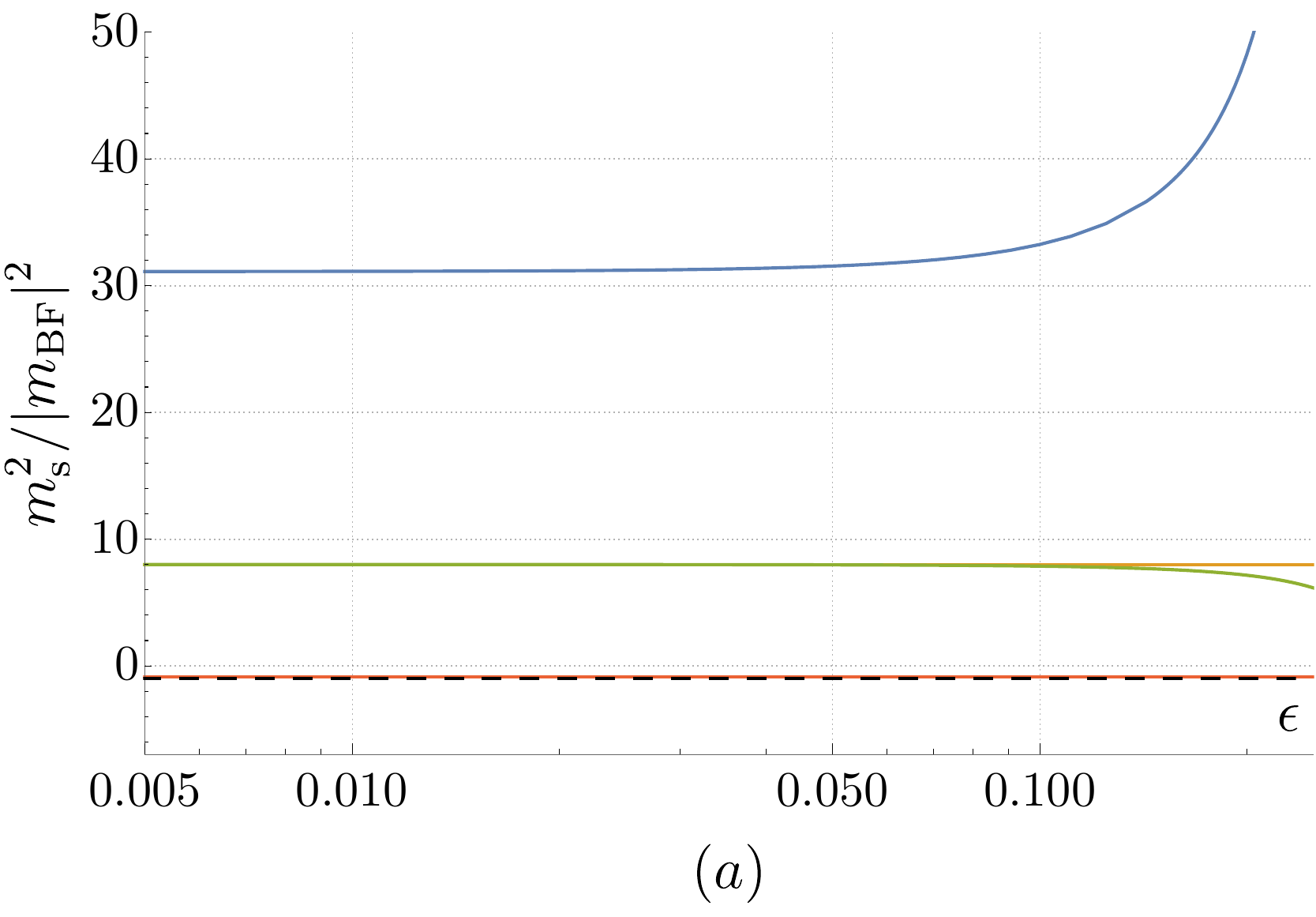}
\quad
\includegraphics[scale=0.27]{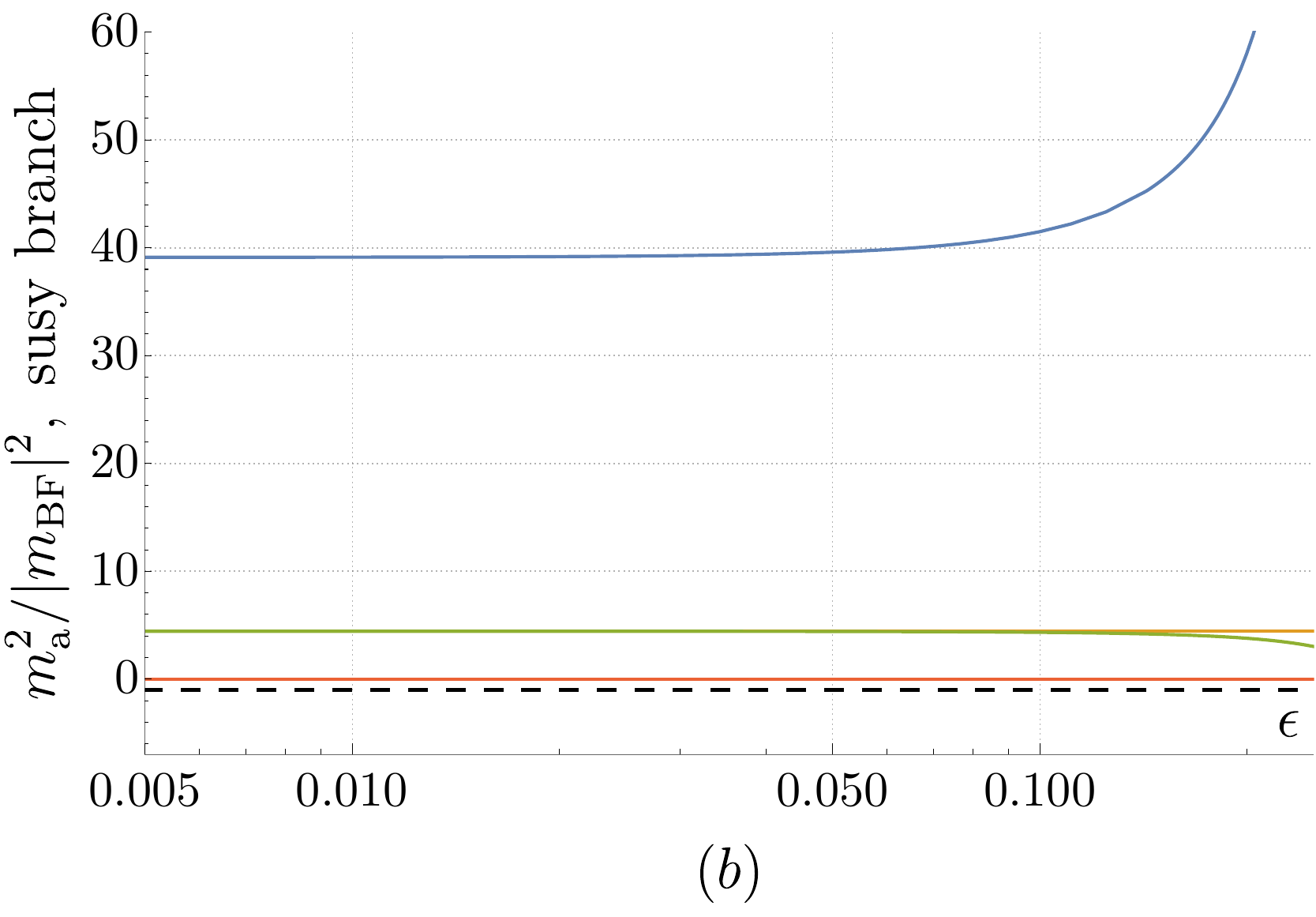}

\includegraphics[scale=0.27]{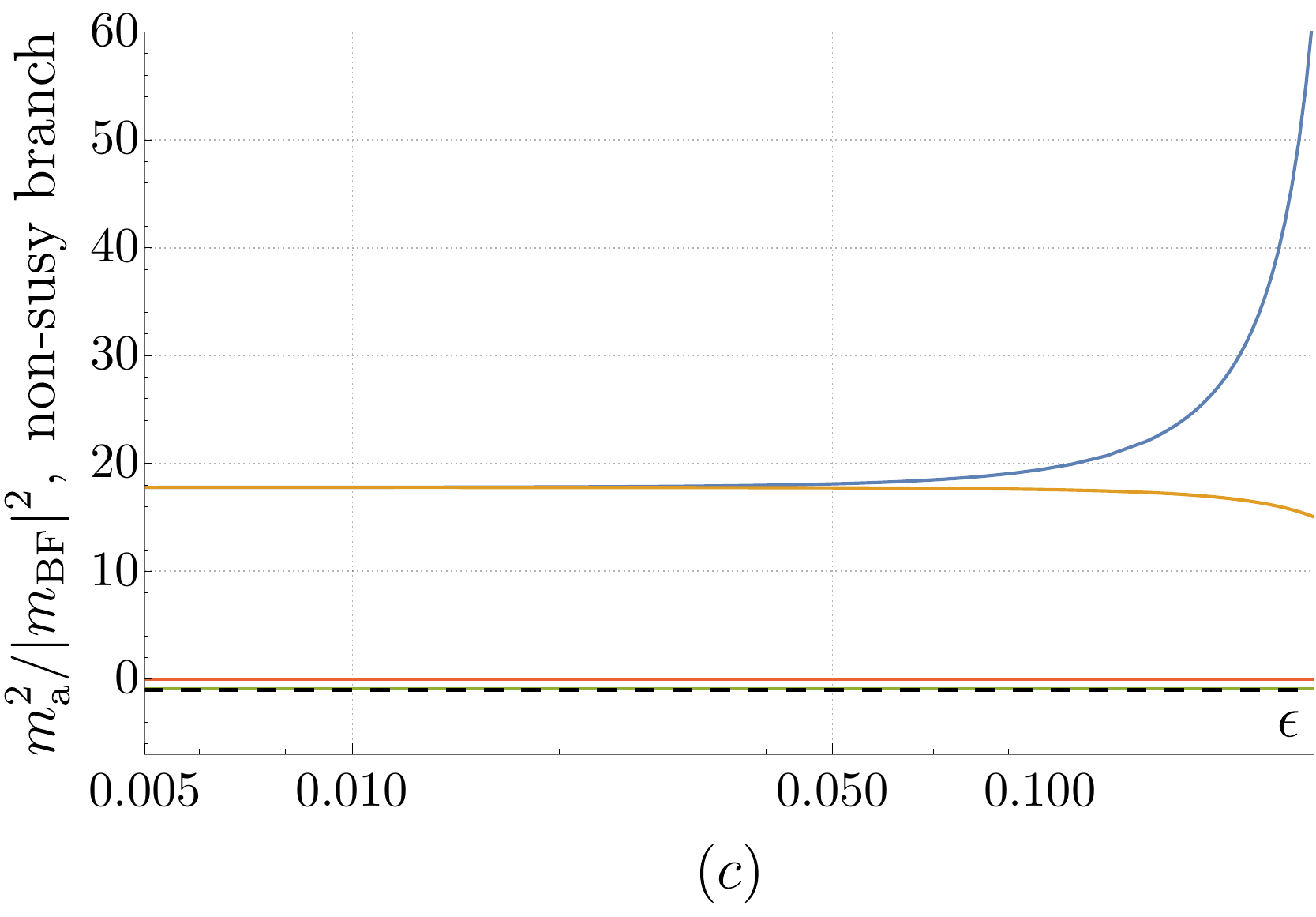}
\caption{The squared masses normalised by the BF bound depicted with the black dashed line for the two-parameter $\mathbb{P}^4_{(1,1,1,6,9)}$ model. (a) The saxionic masses, identical in both supersymmetric and non-supersymmetric branches. The tachyon has degeneracy $h^{2,1}$ and is slightly above the BF bound. (b) The axionic masses in the supersymmetric branch. It is now the massless mode that has degeneracy $h^{2,1}$. (c) The axionic masses in the non-supersymmetric branch. In this family, a single tachyon arises in this sector, with same mass as the saxionic tachyons.}
\label{fig:masses1}
\end{center}
\end{figure}

We then evaluate the scalar masses from the eigenvalues of the Hessian. The saxionic squared masses $m_{\rm s}^2$ normalised by the BF bound  in Planck units $m_{\rm BF}^2\equiv -3/4|V|$ are shown in fig.~\ref{fig:masses1}(a) with respect to $\epsilon=1/n$ on the $x$-axis. These masses are identical in both the supersymmetric and non-supersymmetric branches of vacua. We observe that at small $\epsilon$ the masses converge towards the values derived in the factorised torus or trivial $K3\times T^2$ fibration cases, while being above the BF bound for the whole investigated $\epsilon$ range. Figs.~\ref{fig:masses1}(b) and \ref{fig:masses1}(c) display the normalised axionic masses $m_{\rm a}^2$ for the supersymmetric and non-supersymmetric branches respectively. Again, these masses converge towards the expected values at small $\epsilon$ and do not introduce any instability.

\subsection[The $\mathbb{P}^5_{(1,1,2,8,12)}$ model]{\bm The $\mathbb{P}^5_{(1,1,2,8,12)}$ model}

This three-parameter elliptically fibered geometry has a trivialisation intersection matrix $\eta_{ij}$ and a first Chern class of the base $c^A$, as defined in the core of the paper, see \eqref{kappafib}, given by
\begin{equation}
\eta_{AB}=\begin{pmatrix}
0 & 1\\
1 & 2
\end{pmatrix},
\qquad\qquad
c^A=(0,2)\ .
\end{equation}
Denoting with the index $L$ the direction along the fibre and with $A,B\in\{1,2\}$ the two others, the triple intersection numbers read
\begin{equation}
\cK_{LLL}=8\ ,\quad \cK_{LL1}=2\ ,\quad \cK_{LL2}=4\ ,\quad \cK_{L12}=1\ ,\quad \cK_{L22}=2\ .
\end{equation}

\begin{figure}[ht!]
\begin{center}
\includegraphics[scale=0.27]{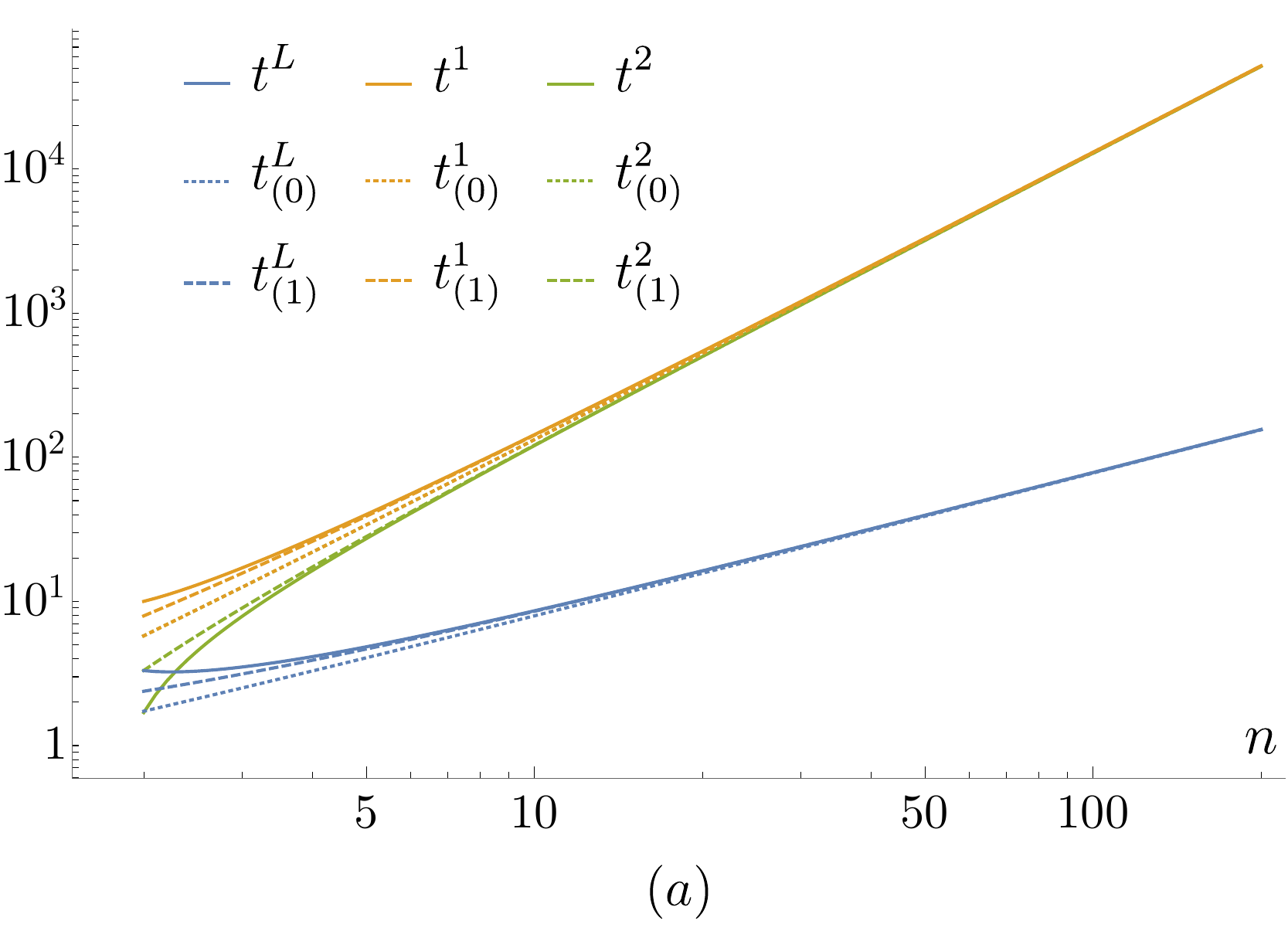}
\quad
\includegraphics[scale=0.27]{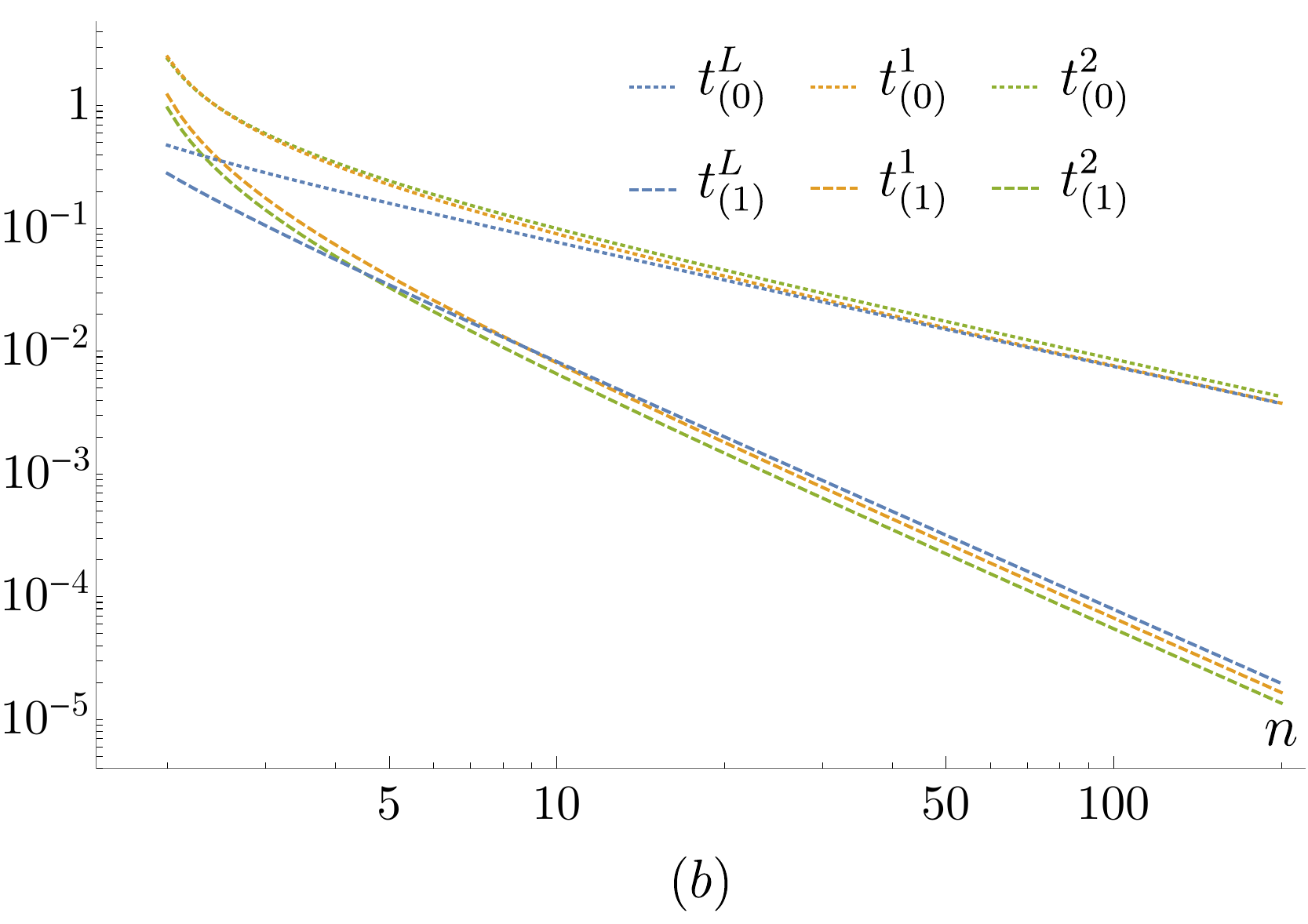}
\caption{(a) Vacuum expectation values of the saxions $t^L$ $t^1$ and $t^2$ for the three-parameter $\mathbb{P}^4_{(1,1,2,8,12)}$ model and their leading-order and next-to-leading order approximations. The green dotted curve coincides with the orange dotted one while $t^1$ and $t^2$ converge to the same asymptotic value because of our choice to have the same unit coefficient in front of the scalings for $m^1$ and $m^2$. That would not be the case with other scalings. (b) The relative error of these approximations with respect to the actual solution. As in the two-parameter case, the plots have been generated in the supersymmetric branch but similar behaviours are observed for non-supersymmetric vacua.}
\label{fig:vevs2}
\end{center}
\end{figure}

Similarly to the two-parameter model studied above, fig.~\ref{fig:vevs2}(a) shows the exact solutions $t^L$ and $t^i$ of the vacuum equations \eqref{saxfib} as well as the leading order and next-to-leading order approximations $t^L_{(0)}$, $t^1_{(0)}$, $t^i_{(1)}$ and $t^i_{(1)}$ with respect to $n$ in the range $[2,200]$. We again observe that the leading order expressions are good approximations of the actual solution and that the next-to-leading order ones are better. The relative errors are displayed in fig.~\ref{fig:vevs2}(b) and they again scale as predicted by the perturbative expansion.

\begin{figure}[ht!]
\begin{center}
\includegraphics[scale=0.27]{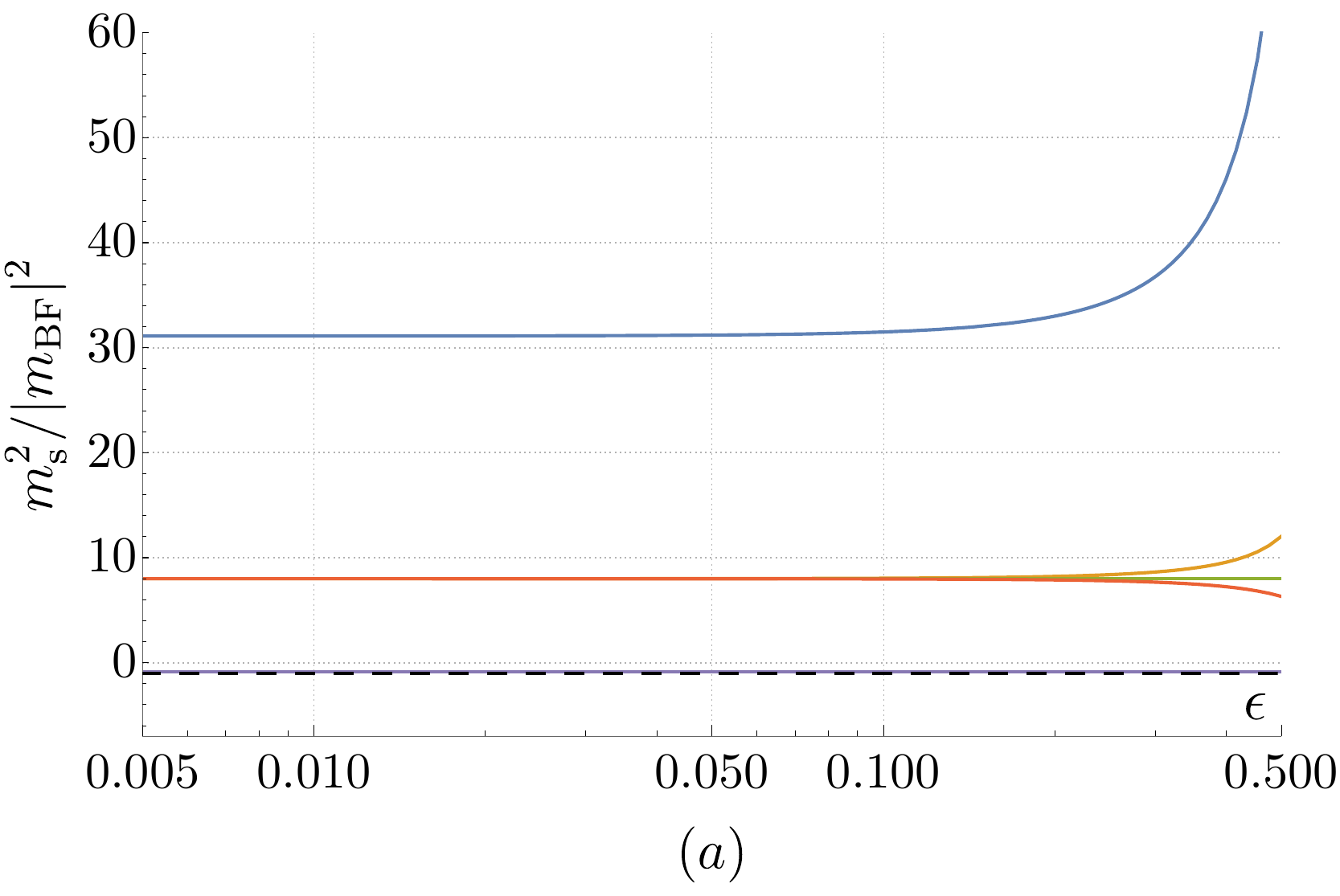}
\quad
\includegraphics[scale=0.27]{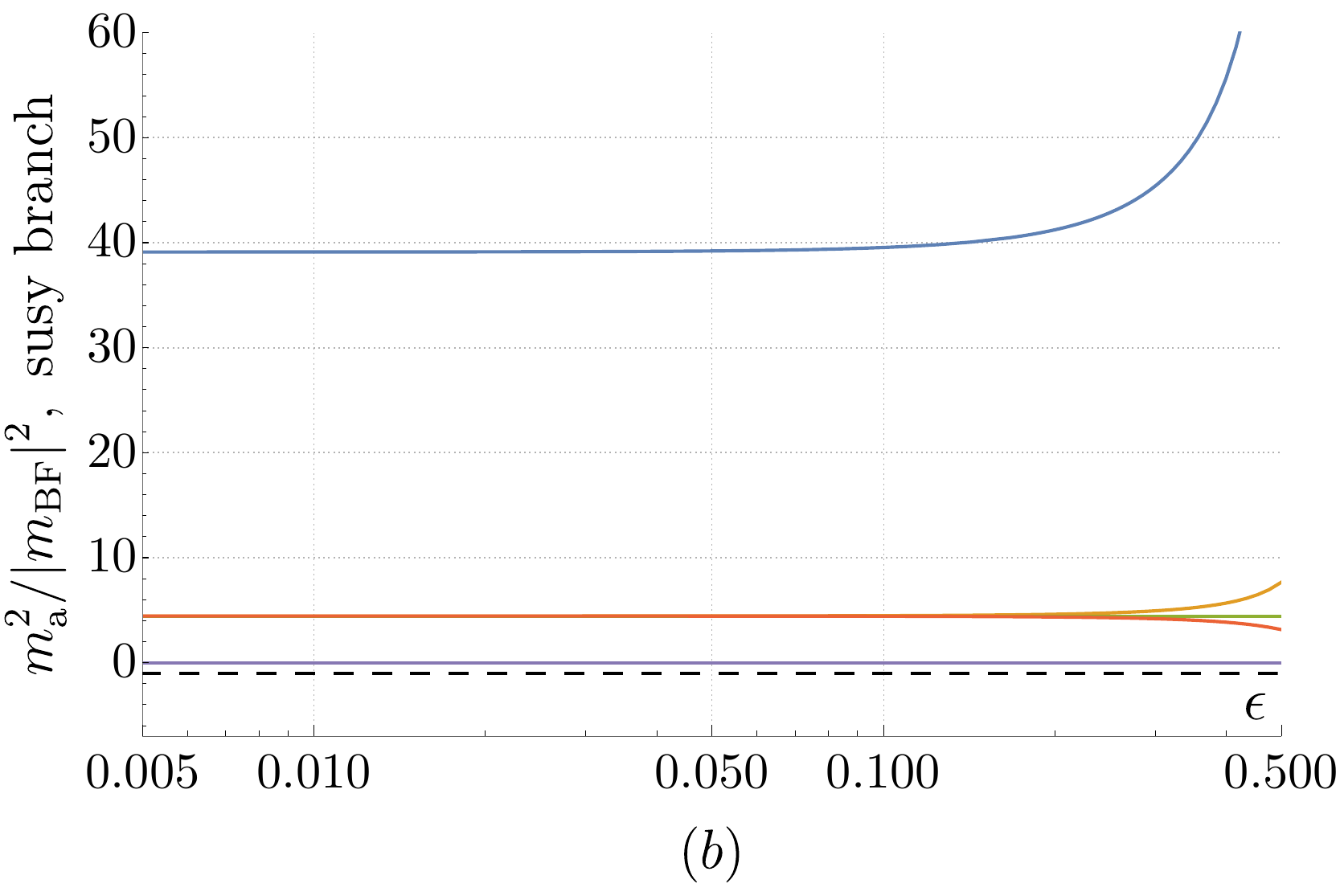}

\includegraphics[scale=0.27]{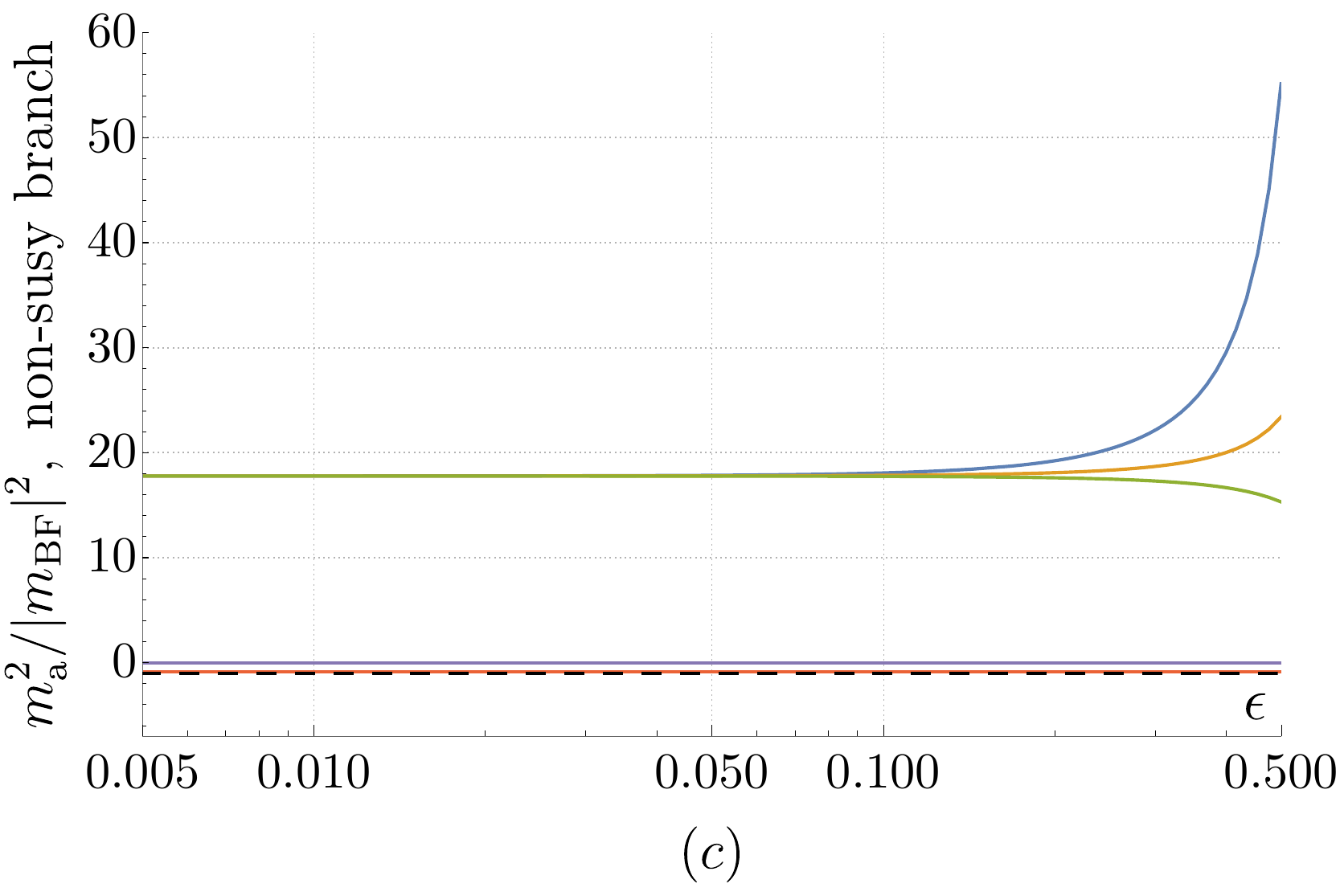}
\caption{The squared masses normalised by the BF bound depicted with the black dashed line for the three-parameter $\mathbb{P}^4_{(1,1,2,8,12)}$ model. (a) The saxionic masses, identical in both supersymmetric and non-supersymmetric branches and with degenerate tachyons slightly above the BF bound. (b) The axionic masses in the SUSY branch. (c) The axionic masses in the non-supersymmetric branch with a tachyon identical to the saxionic one.}
\label{fig:masses2}
\end{center}
\end{figure}

For each vacua, we compute the Hessian from the scalar potential \eqref{eq:potentialgeom} and express it in the primitive/non-primitive basis defined by \eqref{basis}. We compared again our numerical results with a direct implementation of the analytical expressions \eqref{eq:Haxions} and \eqref{eq:Hsaxions} and found again a perfect agreement in this more involved model. This shows the robustness of the analytical derivation of the mass matrix. The saxionic squared masses $m_{\rm s}^2$ normalised by the BF bound defined like in the previous example are shown in fig.~\ref{fig:masses2}(a), while figs.~\ref{fig:masses2}(b) and \ref{fig:masses2}(c) show the axionic masses $m_{\rm a}^2$ for the supersymmetric and non-supersymmetric branches respectively. Again, all these masses converge towards expected values at small $\epsilon$ and do not introduce any instability.

\newpage
\bibliographystyle{JHEP2015}
\bibliography{biblio}

\end{document}